\documentclass{article}
\usepackage{PRIMEarxiv}

\usepackage{listings}
\usepackage{rotating}
\usepackage{amsmath,amsfonts,amsthm,amssymb}
\usepackage{multirow}
\usepackage{pdfpages}
\usepackage[toc,page]{appendix}
\usepackage[section]{placeins}
\usepackage{tablefootnote}
\usepackage{float}
\usepackage{booktabs}
\usepackage{latexsym}
\usepackage{natbib}
\usepackage{bm}
\usepackage{fancybox}
\usepackage[pagebackref=true,colorlinks=true, linkcolor=blue,anchorcolor=red,citecolor=red,filecolor=blue,menucolor=red,urlcolor=black]{hyperref}
\usepackage{array}
\usepackage{geometry}

\newcommand{\be}{{\bf e}}
\newcommand{\bV}{{\bf V}}

\newcommand{\bP}{{\bf P}}

\newcommand{\bQ}{{\bf Q}}

\newcommand{\bA}{{\bf A}}

\newcommand{\bB}{{\bf B}}

\newcommand{\bSigma}{{\boldsymbol{\Sigma}}}
\newcommand{\bLambda}{{\boldsymbol{\Lambda}}}

\newcommand{\bOmega}{{\boldsymbol{\Omega}}}
\newcommand{\bTheta}{{\boldsymbol{\Theta}}}

\newcommand{\bphi}{{\boldsymbol{\phi}}}

\newcommand{\bH}{{\bf H}}
\newcommand{\bD}{{\bf D}}

\newcommand{\bS}{{\bf S}}

\newcommand{\bu}{{\bf u}}

\newcommand{\br}{{\bf r}}
\newcommand{\bh}{{\bf h}}

\newcommand{\bx}{{\bf x}}
\newcommand{\bX}{{\bf X}}
\newcommand{\by}{{\bf y}}
\newcommand{\bY}{{\bf Y}}

\newcommand{\bs}{{\bf s}}

\newcommand{\bff}{\mathbf{f}}

\newcommand{\btau}{{{\boldsymbol{\tau}}}}

\usepackage[utf8]{inputenc} 
\usepackage[T1]{fontenc}    
\usepackage{url}    
\usepackage{booktabs}      
\usepackage{amsfonts}      
\usepackage{nicefrac}       
\usepackage{microtype}      
\usepackage{fancyhdr}       
\usepackage{amsmath}          
\usepackage{graphicx}          
\usepackage{xcolor}
\usepackage{natbib}
\usepackage{amsthm}           
\usepackage{mathtools}
\usepackage{lastpage}          
\usepackage{enumerate}         
\usepackage{parskip}          
\usepackage{caption}
\usepackage{listings}          
\usepackage{algorithm}
\usepackage{algorithmic}
\newcommand{\commentout}[1]{}

\theoremstyle{definition}

\theoremstyle{remark}

\title{Multivariate sensitivity analysis for a large-scale climate impact and adaptation model}

\author{Oluwole. K.~Oyebamiji \\
         Flood and Water Management, 
         HR Wallingford, \\
Howbery Park, 
Wallingford, UK. \\
\texttt{o.oyebamiji@hrwallingford.com} \\
\And
Christopher Nemeth \\
Department of Mathematics \& Statistics, \\ Lancaster University, United Kingdom. \\
\And
Paula A.~Harrison \\
UK Centre for Ecology \& Hydrology, \\
Lancaster, United Kingdom.\\
\And
R. W. ~Dunford \\
UK Centre for Ecology \& Hydrology, \\ Wallingford, United Kingdom.
\And 
G. Cojocaru \\
TIAMASG Foundation, Romania.}

\begin{document}
\maketitle

\begin{abstract}
We develop a new efficient methodology for Bayesian global sensitivity analysis for large-scale multivariate data. The focus is on computationally demanding models with correlated variables. A multivariate Gaussian process is used as a surrogate model to replace the expensive computer model. To improve the computational efficiency and performance of the model, compactly supported correlation functions are used. The goal is to generate sparse matrices, which give crucial advantages when dealing with large datasets, where we use cross-validation to determine the optimal degree of sparsity. This method was combined with a robust adaptive Metropolis algorithm coupled with a parallel implementation to speed up the convergence to the target distribution.  The method was applied to a multivariate dataset from the IMPRESSIONS Integrated Assessment Platform (IAP2), an extension of the
CLIMSAVE IAP, which has been widely applied in climate change impact, adaptation and vulnerability assessments. Our empirical results on synthetic and IAP2 data show that the proposed methods are efficient and accurate for global sensitivity analysis of complex models.

\end{abstract}

\section{Introduction}
There is a growing need in the environmental science community to provide robust decision-making tools for policymakers. By linking together process-based models across sectors, environmental scientists can accurately project future climate change impacts based on climate, social, economic and political factors. Such integrated and cross-sectoral climate change impact models are essential in developing successful evidence-based adaptation strategies \citep{harrison2016climate}. Full uncertainty quantification for these models is critical to ensure that policymakers develop robust strategies to adapt to climate change.
Climate change impact and adaptation models, such as the IMPRESSIONS Integrated Assessment Platform (IAP2) models, are high-dimensional and computationally intensive to evaluate over the full parameter space. To develop a deeper understanding of the key processes which drive these models, it is necessary to identify the relative contributions of the input variables and how they affect the outputs. Many input variables have no significant effect on the output variance, meaning that we can reduce the dimension of the problem by identifying and removing the uncertainties related to these variables, which have little to no impact on the output.

Using sensitivity analysis \citet{gamboa2020global,saltelli2019so}, we can attribute the uncertainty in the outputs of a model to the uncertainty in the model inputs. The principal goal of sensitivity analysis is to identify the crucial parameters that are responsible for most of the variation in the output. Sensitivity analysis can also help identify regions of the input variability space that produce large and extreme output values. It can also be used to identify interactions between variables and highlight where a model can be improved to obtain better input values \citep{farah2014bayesianb,pd5}.

There are two standard approaches for performing sensitivity analysis, namely \emph{local} and \emph{global} sensitivity. Local sensitivity quantifies the variation of the model outputs due to small changes in the uncertain model inputs, where methods such as finite differences and partial derivatives are used as sensitivity indices. We focus our attention on global sensitivity analysis, which shows how the output changes as all the input factors vary continuously over their entire ranges. It uses the full ranges of the uncertainty of the inputs, and these are usually characterised through their joint probability density functions. Global sensitivity requires more computational work than local methods but is capable of revealing hidden relationships between multiple parameters. Computing global sensitivity indices requires the evaluation of a large-dimensional integral over the entire input space. The more parameters we have, the more computationally demanding. Standard numerical integration \citep{robert2004monte,south2020semi}, such as Monte Carlo integration, is usually infeasible when the computer models generating the required model outputs are computationally expensive \citep{farah2014bayesianb}.

Most of the existing global sensitivity analysis techniques assume a univariate output. There exists a large literature that performs global sensitivity analysis of scalar outputs \citep{farah2014bayesianb,svenson2014estimating}. The standard approach used for multivariate outputs is to perform the global sensitivity analysis for each output. The main limitation of this approach, apart from neglecting the output correlation when performing the analysis, is that it produces many sensitivity indices that are difficult to interpret. In contrast to these methods, \citet{gamboa2013sensitivity,lamboni2011multivariate} and recently \citet{xiao2018multivariate,cheng2019multivariate} generalise the Sobol sensitivity indices \citep{sobol1993sensitivity} to the case of multivariate outputs.

\begin{table}
\begin{center}
    \caption{List of user-specific input variables from 2020s to 2080s and their definitions. See Table~\ref{ref1} in Appendix A for the parameter ranges, where $^*$ denotes discrete variables.\label{ref2}}
\fbox{%
\begin{tabular}{*{3}{c}}
\hline
$\bX$& Parameters&Definitions\\
\hline
X$^{\star}_1$ & Emission &Emission scenarios \\
X$^{\star}_2$ & Model & Climate model \\
X$^{\star}_3$ & SSPs & Socio-economic scenarios\\
X$_4$ & Population & Population change\\
X$_5$ & GreenRed & Preference for living close to urban or rural areas \\
X$_6$ & RumLivDem & Change in dietary preference for beef/lamb \\
X$_7$ & NonRumLivDem & Change in dietary preference for chicken/pork \\
X$_8$ & StructChange & Water savings due to behavioural change \\
X$_9$ & TechFactor & Change in agricultural mechanisation \\
X$_{10}$ & TechChange & Water savings due to technological change \\
X$_{11}$ & YieldFactor & Change in agricultural yields \\
X$_{12}$ & IrrigEffFactor & Change in irrigation efficiency \\
X$_{13}$ & GDP & GDP change \\
X$_{14}$ & CostsFactor & Change in oil price \\
X$_{15}$ & ImportFactor & Change in food imports \\
X$_{16}$ & BioEneCropDemand & Change in bioenergy production \\
X$_{17}$ & ArableConservLand & Set-aside \\
X$_{18}$ & CropInpFactor & Reducing diffuse pollution from agriculture \\
X$_{19}$ & Coastal & Coastal flood event \\
X$_{20}$ & Fluvial & Fluvial flood event \\
X$_{21}$ & HabitatRecreation & Protected area changed \\
X$_{22}$ & PA Forestry & Change in protected area forest \\
X$_{23}$ & PA Agriculture & Change in protected area agriculture \\
X$^{\star}_{24}$ & FloodProtection & Level of flood protection\\
\hline
\end{tabular}}
\end{center}
\end{table}

Interpretation of sensitivity indices for models with multivariate outputs is complicated because existing techniques usually ignore the effect of the dimension and correlation between multiple outputs. Dependencies between input variables also complicate this problem if it exists. This paper addresses these shortcomings. Firstly, we develop a multivariate sparse Gaussian process (MSGP) as an efficient, alternative model to replace the expensive computer models. This combines the methods proposed in \citet{svenson2014estimating} and \citet{farah2014bayesianb}. The MSGP model incorporates compact approximations of the dense covariance matrix to reduce the computational burden. The incorporation of sparse matrices offers great savings in storage and computation by introducing some zeros in the matrix of the linear systems.
Secondly, we parallelised the posterior computation by distributing the likelihood estimation across multiple computing nodes using a multicore computing environment. We also exploit high-performance computing libraries such as \texttt{Armadillo} \citep{sanderson2016armadillo}, which is a templated $C^{++}$ linear algebra library to facilitate expensive matrix-matrix multiplication and inversion \citep{eddelbuettel2014rcpparmadillo}. This procedure is combined with a robust adaptive Metropolis-within-Gibbs algorithm \citep{vihola2012robust} to efficiently sample from the posterior target distribution.
Thirdly, we extended the computation of sensitivity indices to the case of multivariate spatially distributed outputs for computationally demanding models. This integrated approach produces a reduced-form representation of the model that enabled a substantial reduction in computation time.

The paper is organised as follows. Section 2 provides a brief description of the IMPRESSIONS IAP2 model and parameter sampling. The multivariate sparse Gaussian processes (MSGP) theory, its parameter estimation, and compact correlation functions are given in Section 3. Section 4 focuses on the proposed new sensitivity indices, while the analysis and results from the synthetic and IAP2 data are presented in Section 5. The paper closes with a conclusion in Section 6.

\section{Model and simulation data}
\subsection{IMPRESSIONS Integrated Assessment Platform version 2 (IAP2) model}
The IMPRESSIONS IAP2 model was developed to explore the impacts of climatic and socio-economic change on the interactions between different landscape sectors (agriculture, forestry, water resources, biodiversity, coastal and fluvial flooding, and urban development) \citet{harrison2019differences}. It provides an integrated approach to support climate change adaptation decision-making by helping decision-makers understand the complex interactions between sectors and how these are affected by changing climatic and socio-economic conditions. The latest scenarios from the Intergovernmental Panel on Climate Change research community are incorporated into the IAP2 for climate (based on Representative Concentration Pathways) and socio-economics (based on the Shared Socioeconomic Pathways). The model produces outputs for a wide range of both sector-based impact and vulnerability indicators and ecosystem service indicators to link impacts directly to human well-being \citep{harrison2016climate}.

The IMPRESSIONS IAP2 operates at a spatial resolution of 10 $\times$ 10 arcminutes across Europe. The models within the IAP2 are hard-linked, with flows of data passed between models as part of a hierarchical model chain. Each of the ten models within the IAP2 is represented by a simplified meta-model to enable the full model chain to run with a web-based interactive interface \citep{harrison2015assessing,kebede2015direct}. Nevertheless, a single run of the model takes some minutes, and multiple runs can not be made parallel because it is a graphical-based model. The IMPRESSIONS IAP2 model has a large number of input parameters, namely climate variables, such as precipitation and temperature, and socio-economic variables, such as the change in the population, GDP, agricultural mechanisation and diets. Although the IAP2 adopts a meta-modelling approach for its ten component models, the expense of running such a model means that it is infeasible to run all possible parameter value combinations.

The model produces over 100 outputs, which can be categorised into land use-related indicators, such as the area of urban land, managed forests and arable land, and other indicators related to food production, carbon storage, species suitability, water exploitation and flood risk. In addition to the baseline, it simulates impacts for three timeslices (the 2020s, 2050s and 2080s), three emissions scenarios (RCPs 2.6, 4.5 and 8.5) and four socio-economic scenarios (SSPs 1, 3, 4 and 5). The ten climate models used under various emissions scenarios are given in Appendix \ref{appA}. Further details about the model and scenarios are given in \citet{harrison2019differences}.

\subsection{Parameter sampling}
Sampling the input parameter space is a critical part of both sensitivity analysis and building a useful surrogate model. The goal is to explore the entire input space with a reasonable sample size $n$. There exist a large number of various sampling schemes \citep{pd5,burhenne2011sampling}.

A proper sample generation technique is crucial for efficient surrogate modelling for computationally-expensive problems. It ensures an accurate estimation of model parameters. This enables large-scale sensitivity analyses and scenario exploration without the computational burden. Moreover, most of the work on sensitivity and uncertainty analyses from large-scale models assume that the input factors in computer models are continuous variables, and most of the existing methods are not directly applicable to applications that contain both continuous and categorical variables. The presence of categorical variables makes the design of computer experiments a more laborious task because the qualitative variables do not allow for an infinite number of different points or levels.

The input parameters for running a set of runs from the IMPRESSIONS IAP2 model were generated using Latin Hypercube Sampling (LHS) \citep{pd5}. The procedure provides data for training a robust surrogate model to approximate the outputs from the IMPRESSIONS IAP2 model. The LHS technique is popular and provides better coverage of the input space than standard random sampling techniques. We used the optimum LHS technique to avoid clustering of the sample points and to ensure a good estimation of the statistical moments of response functions. Optimum LHS also maximises the mean distance from each design point to all the other points such that points are as spread out as possible \citep{carnell2019package,stocki2005method}.

We selected 24 input parameters to evaluate the sensitivity of the IMPRESSIONS IAP2 model outputs; 20 of these variables are continuous parameters (see Table~\ref{ref2} of Appendix \ref{appA}). The four categorical variables are level of flood protection, emission, climate model and socio-economic scenario (denoted as $^{\star}$ in Table~\ref{ref2}). We used a two-stage sampling technique. In the first stage, we use the LHS to sample the continuous variables, and in the second stage, we formed every possible combination of categorical variables according to the range of values given in Table~\ref{ref1}. We limit our samples to just $2,000$ design points. This gives $\sim 30,000$ design points in total (Note: 30,000 simulations takes $\sim$ 42 days of computer time), and each sampling point produces over a hundred spatially-distributed outputs.

\subsection{Output data}
\begin{figure}[H]
\begin{center}
\begin{minipage}{.8\textwidth}
\centering
\includegraphics[width=.9\linewidth]{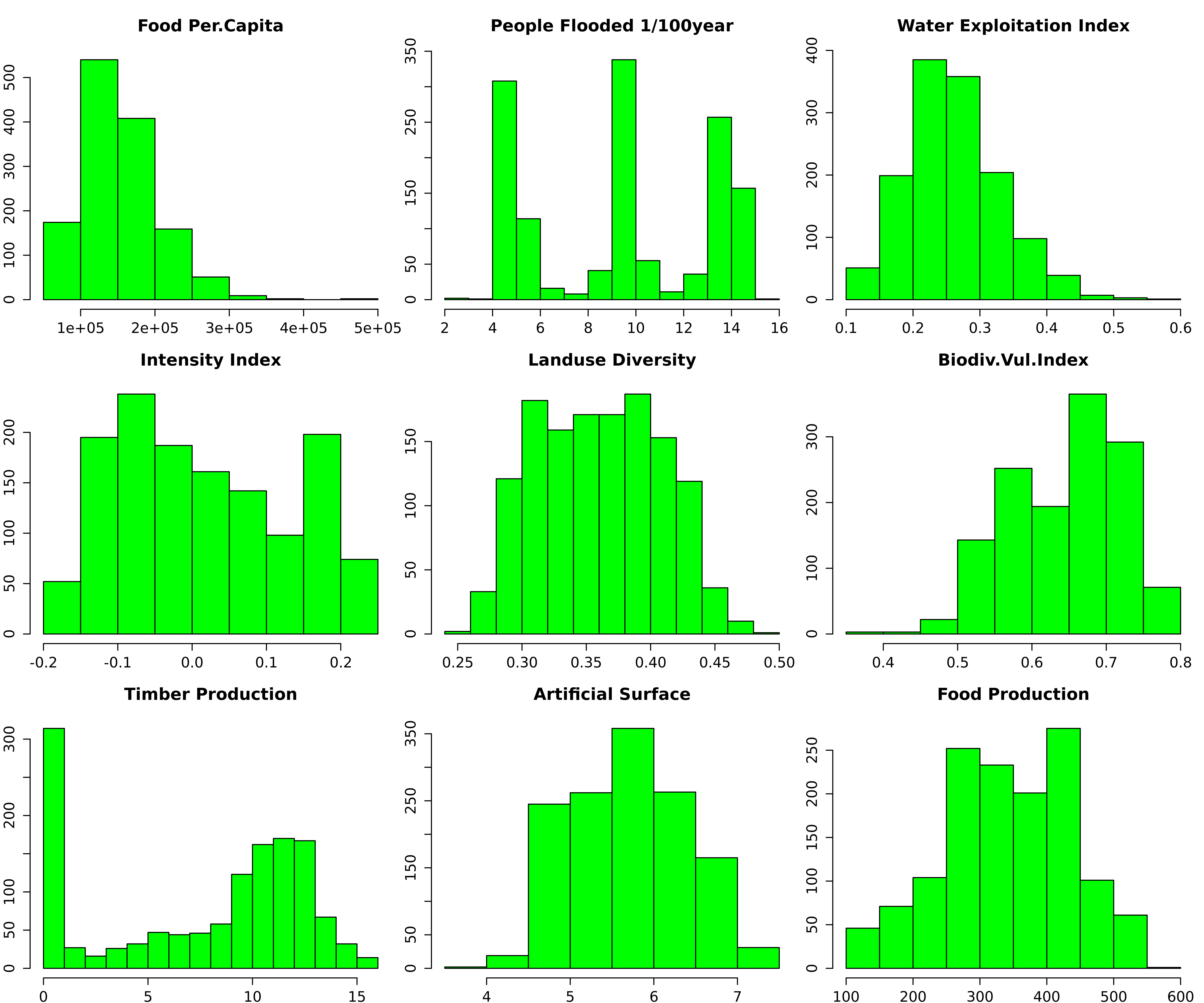}
\end{minipage}
\\
\begin{minipage}{.8\textwidth}
\centering
\includegraphics[width=.9\linewidth]{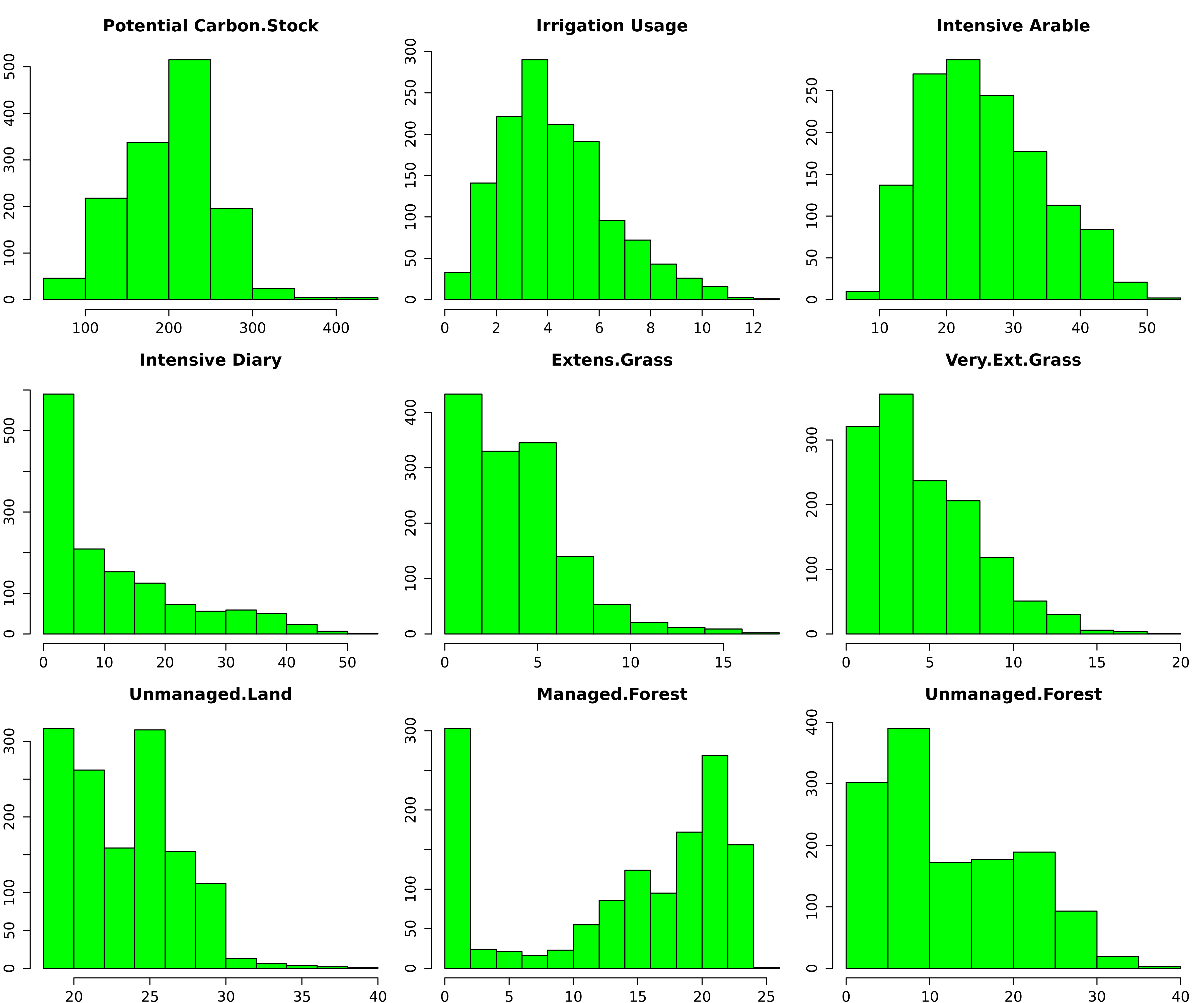}
\end{minipage}
\caption{Histograms of spatially aggregated IAP2 simulation for the 18 selected outputs, averaged across the three emission scenarios and ten climate models for SSP1.}\label{fig2}
\end{center}
\end{figure}

We used the LHS algorithm to produce ensembles of simulations from the IMPRESSIONS IAP2 model.
We have selected 18 important variables that reflect impacts across sectors (see Table~\ref{ref3}). The data are on a $10^{\circ} \times 10^{\circ}$ resolution grid $\sim 16 km^2$ and each has 23,871 grid cells. Figure~\ref{fig2} shows the histograms of selected IAP2 simulation outputs under the SSP1 scenario (which is characterized by a sustainable future with global cooperation and less inequality), averaged over all emissions scenarios, climate models and timeslices.
The histograms of \textit{food per capita} and \textit{water exploitation index} are relatively similar and symmetric, while that of \textit{people flooded in a 1 in 100-year event} has three major peaks. The histogram of \textit{timber production} and \textit{managed forest} also have a single major peak and a minor peak, making them bimodal distributions. The remaining variables such as \textit{irrigation usage, intensive grassland arable, intensive grassland dairy, extensive grassland} and \textit{very extensive grassland} are right-skewed. Apart from the nonsymmetric nature of most of the variables with no distinct shape, we note that the range of values of these variables varies widely. The diverse structural patterns clearly show the complexity of the effects we are modelling. Our model formulation needs to capture these divergent behaviours to be useful for performing sensitivity analyses. The histograms of the remaining simulated IAP2 outputs for SSP4 are given in Figure~\ref{fig2b} in Appendix \ref{appA}.

\begin{table}
\caption{Selected outputs. \label{ref3}}
\centering
\fbox{
\begin{tabular}{*{2}{c}}
\hline
Variables &  \\
\hline\hline
Food per capita & People flooded in a 1 in 100 year event \\
Water exploitation index & Landuse intensity index \\
Land use diversity & Biodiversity vulnerability index\\ Timber production & Area of artificial surfaces\\
Food production & Potential carbon stock \\ Irrigation usage & Arable land \\
Intensive grassland & Extensive grassland \\ Very extensive grassland & Unmanaged land \\
Managed forest & Unmanaged forest
\end{tabular}}
\end{table}

\section{Surrogate Modelling}
\subsection{Multivariate sparse Gaussian processes (MSGP)}\label{senn1}
We introduce the overall statistical model structure in this section and describe each component of the model. We formulate a probabilistic model for multivariate data using a multivariate Gaussian process and describe a Markov chain Monte Carlo (MCMC) sampling algorithm for inference. We also outline how to extend this idea to sparse approximation models for efficient sampling and achieve dimension reduction for large multi-dimensional data. The main procedure involves incorporating a sparse approximation to achieve efficient matrix operations within a Bayesian framework. This multilevel model is estimated using an adaptive MCMC technique within a parallel computing environment. We implement an adaptive Metropolis algorithm \citet{vihola2012robust} within a Gibbs sampler to facilitate highly efficient posterior sampling.

Let $\mathbf{x}=\left(x_{1}, \ldots, x_{p}\right)^{\mathrm{T}} \in \mathcal{X} \subset \mathbb{R}^{p}$ be the vector of $p$ input variables with outputs from the simulator $f: \mathbf{x} \rightarrow \mathbf{y} \in \mathcal{Y} \subset \mathbb{R}^{m}$ denoted as $\mathbf{y}$, i.e. 
$$f(\mathbf{x})=\left\{f_{1}(\mathbf{x}), \ldots, f_{m}(\mathbf{x})\right\}^{\mathrm{T}}$$
is the $m \times 1$ output vector from the simulator at input combination $\bx$. Also, let the matrix of input combinations be $\bX=\big(\mathbf{x}_{1}, \ldots, \mathbf{x}_{n}\big),$ with $\mathbf{x}_{i}=\left(x_{i 1}, \ldots, x_{i p}\right)^{\mathrm{T}}$ representing the $i^{th}$ column. The multivariate outputs from the
simulator, with $n$ sample points, are combined into an $n \times m$ output matrix
$
\bY=\left(\begin{array}{c}
f\left(\mathbf{x}_{1}\right)^{\mathrm{T}} \\
\vdots \\
f\left(\mathbf{x}_{n}\right)^{\mathrm{T}}
\end{array}\right).
$
We can decompose $\bY$ into a linear trend and nonlinear random process,
\begin{equation}\label{eq1a}
\begin{array}{l}
\bY = \bH(\bx)\boldsymbol{B}+ \boldsymbol{\varepsilon}(\bx),
\end{array}
\end{equation}
where $\bB$ is a $p\times m$ matrix of regression coefficients capturing most of the variation in $f(\cdot)$ and $\bH=\left[{h}\left({\bx}_{1}\right) \cdots {h}\left({\bx}_{n}\right)\right]^{\mathrm{T}}$ is a $n \times p$ regression matrix of explanatory variables $h(\cdot)$, common to all the outputs and $\boldsymbol{\varepsilon}(\cdot)$ is a vector random field, modelled as a multivariate Gaussian process. We use a separable covariance function in the matrix-normal distribution $\boldsymbol{\varepsilon} \sim \operatorname{MN}\left(\mathbf{0}, \boldsymbol{\Sigma}, \mathfrak{R}\right)$ so that we capture the remaining unexplained small scale structure. The $n \times n$ input correlation matrix $\mathfrak{R}$, i.e. 
$\operatorname{Cor}\left(\boldsymbol{\varepsilon}(\mathbf{x}), \boldsymbol{\varepsilon}\left(\mathbf{x}^{\prime}\right)\right)= R\left(\mathbf{x}, \mathbf{x}^{'} \right),$ is chosen by the user. A common choice for $R(\cdot)$ is the exponential power correlation function \citep{kaufman2011efficient}, given in Equation ~\eqref{expp}. Further details regarding the correlation functions used in this analysis are given in the next section. The $m \times m$ system scale matrix $\bSigma$ denotes the cross-covariance matrix between the $m$ multivariate outputs.

To predict at $n^*$ new test values $\bx^*$ with given estimated parameters $\bTheta$, the conditional distribution of $\bY^*$ given training data $\bD$, is a matrix-normal with mean and covariance given respectively as
\begin{equation}\label{md1}
\bY^* | \bx^*, \mathbf{D}, \bB, \bSigma \sim MN(\bQ, \bSigma, \br^*-\br^{T}\mathfrak{R}^{-1}\br),
\end{equation}
where $\bQ = \bH^*{\bB}+\mathbf{r^*}^{T} \mathfrak{R}^{-1}(\mathbf{D}-\bH {\bB})$ and $\mathbf{r}$ is the $n \times n^*$ matrix of cross-correlations between the observed responses $\bX$ and test points $\bx^*$, while $\mathbf{r}^*$ is the $n^* \times n^*$ correlation matrix between $\bx^*$. To achieve further efficiency, we can integrate out  $\bB$ and $\bSigma$ to give
\begin{equation}\label{md2}
\bY^* | \bx^*, \bD \sim MT(\bQ, \bS,\mathfrak{D}, \delta),
\end{equation}
where $MT(\cdot)$ is the matrix t-distribution with location parameter $\bQ$
and a row-scale matrix
\begin{equation}\label{pre2}
\mathfrak{D} = \br^*-\br^{T}\mathfrak{R}^{-1}\br + (\bH^* - \br^{T}\mathfrak{R}^{-1}\bH ) {\bLambda} (\bH^* - \br^{T}\mathfrak{R}^{-1}\bH )^{T}
\end{equation}
while $\bS$ and $\delta$ are column-scale matrices and degrees of freedom, respectively.

\subsection{Compactly supported correlation functions}\label{compact}
We note that making inference for the model parameters, and creating predictions from the model, is computationally expensive. The main bottleneck is in the computation of the inverse and determinant of an $n\times n$ matrix $\mathfrak{R}$ and solving for $\mathfrak{R}^{-1}(\bY - \bH \bB)$. These operations scale with an order of $\mathcal{O}(n^3)$ and are thus intractable for large $n$, where $n$ is the sample size.
To overcome this problem, we introduce compactly supported correlation functions to multivariate outputs in this study. These correlation functions have been used in earlier works of \citet{kaufman2011efficient} and \citet{svenson2014estimating}, which focus only on univariate applications.

The concept of compactly supported correlation functions was proposed by \citet{gneiting2002compactly} to parameterise the smoothness of the associated stationary and isotropic random field. A stationary process that employs standard covariance functions for its predictions produces dense covariance matrices, and this is problematic, especially when $n$ is large. Compactly supported correlation functions yield sparse covariance matrices. Having many zero entries in the covariance matrices can both greatly reduce computer storage requirements and the number of floating-point operations needed in computation. These types of functions with compact support are used to facilitate efficient computation in spatial predictions. A comprehensive review of compactly supported covariance functions is given in \citet{moreaux2008compactly,gneiting2002compactly}.

Suppose $R\left(\mathbf{x}, \mathbf{x}^{'}; \boldsymbol{\phi}\right)$ is a power exponential correlation function which is a function of the Euclidean distance between $\bx$ and $\bx'$, we have
\begin{equation}\label{expp}
R(\bx,\bx')=\prod\limits_{k=1}^{p} \exp \left\{-\bphi_{k}\left|x_{k}-x_{k}^{\prime}\right|^{2}\right\} \quad \text{(Power exponential)},
\end{equation}
where $\bphi$ is a correlation range parameter for a finite set of points.
The idea is that for many data sets, the correlations vanish beyond a certain cut-off distance.
We have used two functions, given below, that can be directly used to approximate the power exponential function. Therefore, $\phi_k$ is no longer used as a parameter in the model and instead we have a cut-off parameter $\tau_{k} \geq 0$ and $t=|x-x'|.$
$$\text { Bohman: } \quad R(t ; \tau)=(1-t / \tau) \cos (\pi t / \tau)+\sin (\pi t / \tau) / \pi, 0\leq t/ \tau \leq 1,$$
$$\text { Truncated power: } \quad R(t ; \tau, \alpha, v)=\left[1-(t / \tau)^{\alpha}\right]^{\nu}, ~ 0<\alpha<2, \nu \geq \nu_{p}(\alpha).$$

To ensure positive-definiteness for the truncated power function, there is a constraint on the parameter  $\nu_p(\alpha)$ so that it will produce a valid correlation matrix, $\lim_{\alpha \rightarrow 2} \nu_p (\alpha) =\infty$. The range of upper bounds for several values of $\alpha$ between 1.5 and 1.955 were given in \citet{gneiting2001criteria}
because it is difficult to evaluate $\nu_p(\alpha)$. For instance, the upper bounds when $\alpha = \{3/2, 5/3\}$ are 2 and 3, respectively.

We also need to restrict $\tau$ to fall within the p-dimensional triangle to impose sparsity in the correlation functions above, i.e. $$\tau_{c}=\left\{\tau \in \mathbb{R}^{p}: \tau_{k} \geq 0, \quad \sum_{k=1}^{p} \tau_{k} \leq c\right\}, \quad \forall k, c>0.$$
To achieve some levels of sparsity, parameter $c$ is fixed to a reasonable value such that a given percentage of the $\omega$ off-diagonal elements of $\mathfrak{R}$ are zeroes. The procedure for determining an appropriate $c$ to obtain a required $\omega \%$ of zeroes among the off-diagonal elements is given below. Compute a pair distance $$d_{i, j} = \sum \limits_{k=1}^{p}\left|\bx_{i k} - \bx_{jk}\right| \ \  \mbox{for} \  1 \leq i<j \leq n.$$ Let $c$ be the $\left\lfloor\left(\begin{array}{c}n \\ 2\end{array}\right) \times \omega \%\right\rfloor^{th}$ smallest value among the $d_{i,j}$ 's, where $\lfloor\cdot\rfloor$ is the integer part of $\left(\begin{array}{c}n \\ 2\end{array}\right) \times \omega \%$. We note that the cutoff $\tau_{c}$ controls both the correlation range and degree of sparsity. See  \citet{svenson2014estimating} and \citet{kaufman2011efficient} for further details.


We have allowed for various degrees of smoothness in our analysis. For instance, the Matern correlation corresponds to a process which is $r$ times mean-squared differentiable if $\nu>r$. Unlike the Matern correlation, the truncated power function is not differentiable at the origin, even with $\alpha < 2$, corresponding to a process that is not mean-squared continuous. The Bohman function, on the other hand, is twice differentiable at the origin when $\alpha < 2$ and leads to a process which is mean-square differentiable. The power exponential function, eq.\eqref{expp}, is smooth and infinitely differentiable at the origin when $\alpha = 2$.
We note that the compact correlation function preserves both compact support and the local behaviour of the correlation function at the origin. Therefore, compactly supported correlation functions can be directly used to model dependencies within random processes \citep{kaufman2011efficient,svenson2014estimating,gneiting2002compactly}, which is the approach we follow in this paper. They can also be used by multiplying with another strictly positive correlation function, often called tapering, for estimation and prediction in spatial statistics \citep{pd7b,kaufman2008covariance}. Further details are given in Appendix \ref{appD} and a summary of the implementation procedure is given in the MCMC algorithm in the Appendix \ref{appB}.

Other popular alternatives to reduce the computational burden are the earlier works of \citet{vecchia1988estimation,stein2004approximating} where the joint likelihood is partitioned and written as a product of conditional densities. The smaller sets of conditional sets consisting of nearest neighbours are then chosen. This approach was recently developed into a nearest neighbour Gaussian process using parallel computation in the works of \citet{finley2017applying,datta2016nonseparable} and \citet{zhang2019practical}. This approach uses a conditional independence assumption to learn a sparse inverse covariance.

\subsection{Posterior sampling}\label{sec:sampling}

Let $\bTheta=\left\{\bB, \bSigma, \boldsymbol{\btau}\right\}$ be the set of unknown parameters and $p(\boldsymbol{\Theta})$ the prior distributions for these unknown parameters.
Using the Bayesian approach, we can estimate these unknown parameters by assigning a conditionally conjugate prior, $p(\boldsymbol{\bB} | \boldsymbol{\Sigma},\btau)p(\boldsymbol{\Sigma}|\btau)p(\btau)$. We use a matrix-normal inverse-Wishart (MNIW) prior for $\left\{\bB, \bSigma \right\}$, where we assign a normal prior to the matrix of regression coefficients $\boldsymbol{\bB} | \boldsymbol{\Sigma},\btau \sim N\left(\boldsymbol{\bB}_{0}, \boldsymbol{\Sigma} \otimes \mathbf{\Lambda_{0}}\right)$ and cross-covariance $\boldsymbol{\Sigma}|\btau \sim \mathcal{W}^{-1}\left(\mathbf{\bS}_{0}, \delta_{0}\right),$ where $\bB_0$, $\Lambda_0$, $\bS_0$ and $\delta_0$ are prior hyperparameters. 


We can use MCMC to sample from the posterior $p(\bTheta|\bY)$ using a Metropolis-within-Gibbs algorithm. This algorithm combines Gibbs sampling with an adaptive random-walk algorithm. We use a Gibbs sampler to sample the parameters $\bB$ and $\bSigma$ using their conditional posterior distributions,
\begin{equation}\label{eq2}
\boldsymbol{\bB} | \bY, \bSigma, \btau
\sim \mathcal{M} \mathcal{N}_{n,m}\left({\bB}, \bSigma,{\Lambda}\right) \ \ \mbox{and} \ \ \boldsymbol{\bSigma} | \bY, \bB, \btau
\sim \mathcal{W}^{-1}\left(\bS, \delta\right),
\end{equation}
where $\widehat{\bB}=\widehat{\bLambda} \left(\mathbf{H}^{\mathrm{T}}\mathfrak{R}^{-1} \mathbf{Y}+\mathbf{\Lambda}_{0} \mathbf{B}_{0}\right)$, $\widehat{\bLambda}=\left(\mathbf{H}^{\mathrm{T}}\mathfrak{R}^{-1} \mathbf{H}+\mathbf{\Lambda}_{0}\right)^{-1}$, $\widehat{\bS}=\mathbf{\bS}_{0}+ \mathbf{Y}^{\mathrm{T}}\mathfrak{R}^{-1}\mathbf{Y}
+\bB_0^{\mathrm{T}} \mathbf{\Lambda}_{0}\widehat{\bB_0}
-\widehat{\bB}^{\mathrm{T}} \mathbf{\Lambda}\widehat{\bB}$ and $\hat \delta=\delta_{0}+n_0.$

We sample the parameter $\btau$ using a robust adaptive Metropolis-Hastings algorithm \citep{vihola2012robust}. We can simplify the posterior distribution for $\btau$ by integrating out the matrices $\bB$ and $\bSigma$,
\begin{equation}
p\left(\boldsymbol{\btau}| \bY\right) \propto |\mathfrak{R}|^{-\frac{m}{2}} \times |\widehat{\bLambda}|^{m/2} |\bS|^{-(\delta+m-1)/2} \times p(\btau).
\end{equation}
See \citet{rougier2007lightweight,rougier2008efficient} and \citet{overstall2016multivariate} for further details. The robust adaptive Metropolis algorithm is highly efficient as it automates the tuning of the random-walk proposal. Further details on the full MCMC procedure, including pseudo-code, are given in Appendix \ref{appB}.

\section{Performing sensitivity analyses}
\subsection{Evaluating global sensitivity indices using MSGP}\label{senn2}
The IAP2 model we are using in this paper consists of $p=24$ input variables, which produce multivariate output data, from which we have carefully selected m = 18 responses for this analysis. This would require the estimation of $p\times m$ sensitivity indices. Using the compactly supported correlation functions to model small-scale variability reduces the computational expense as operations involving the zero elements need not be stored or performed. We used a two-stage approach to compute all the sensitivity indices. The first step is to sample from the posterior distribution of $\widehat \bTheta$ using an adaptive Metropolis within Gibbs sampler as described in the previous section. As noted earlier, our choice of prior distribution allows us to work with the marginal distribution. The second step is to compute predictive means and variances for new inputs using the posterior samples $\widehat \bTheta$ and then computing the sensitivity indices.

The variance-based approach uses the popular Hoeffding decomposition method of \citet{van2000asymptotic} to separate the total variance into different partial variances for a univariate output $Y$. These partial variances measure the associated output uncertainty induced by the corresponding input variables. By considering the ratio of each partial variance to the total variance, we can obtain a measure of importance for each input variable called the sensitivity index. Suppose $E\left(Y | x_{j}\right)=\int_{\bx_{\sim j}} f(\mathbf{x}) \mathrm{d} G_{\sim j | j}\left(\mathbf{x}_{\sim j} | x_{j}\right),$
where $\bx_{\sim j}$ denotes input vector $\bx$ excluding element $x_j$,
then the variance-based method of decomposition into partial variances gives the portion of variation $V$ that is explained by the model inputs either independently or in combination with other factors. For a $p-$ dimensional input factor, we have
\begin{align}\label{md3}
Y= f(\bx)&=f_{0}+\sum_{j=1}^{p} f_{j}\left(x_{j}\right) \\
&+\sum \limits_{j_1=1}^{p-1} \sum \limits_{j_2 = j_1+1}^p f_{j_1, j_2}\left(x_{j_1}, x_{j_2}\right)+\cdots  \quad+f_{1,2, \ldots, p}\left(x_{1}, \ldots, x_{p},\right) \\
&=f_{0} + f_{j} + f_{\sim j} + f_{j, \sim j}
\end{align}
where $f_{0}=E(Y)=\int_{\bx} f(\bx) dG(\bx),$ denotes the global mean and $G(\bx)=\prod_{j=1}^{p} G_{j}\left(\bx_{j}\right)$ is the
probability distribution of the input factors by assuming that each factor is independent with components $G_{j}\left(x_{j}\right) $. Similarly, the next $p$ terms $f_{j}\left(x_{j}\right)$ are the main effect of input $x_{j},$ providing a measure of the influence of input $x_{j}$ on the model output, where

$$f_j(x_j) = E(Y|x_j) -E(Y)=\int_{\boldsymbol{x}_{\sim j}} f(\boldsymbol{x}) dG\left(\boldsymbol{x}_{\sim j} | x_{j}\right) - E(Y),
\ \mbox{for} \ j=1, \ldots, p.$$

We also define $\bx_{\sim j}$ to represent input vector $\bx$ excluding element $x_j$ and using the independence assumption between the input factors, then the expression $G(\bx_{\sim j}|x_j)$ can be rewritten as $G(\bx_{\sim j})$. The remaining terms of the decomposition are the interactions, which quantify the combined influence on the output of two or more inputs taken together.

\citet{homma1996importance} and \citet{sobol1993sensitivity} generalise this output decomposition into a summation of partial variances, since each component of eq.~\eqref{md3} is orthogonal such that
\begin{equation}\label{md4}
V=\sum_{j=1}^{p} V_{j}+ \sum \limits_{j_1=1}^{p-1} \sum \limits_{j_2 = j_1+1}^p V_{j_1,j_2}+\cdots+V_{1,2, \ldots, p},
\end{equation}
where $V= \operatorname{var}(Y)$, $V_{j}=\operatorname{var}\left(f_{j}\left(x_{j}\right)\right)=\operatorname{var}\left(E\left(Y | x_{j}\right)\right)$, and $V_{j_1, j_2}=\operatorname{var}\left(f_{j_1, j_2}\right.$
$\left.\left(x_{j_1}, x_{j_2}\right)\right).$ These results hold similarly for the higher order terms.
Standardizing equation \eqref{md4}, produces
$$S_{j}=\frac{V_{j}}{V}, \quad \quad S_{j_1, j_2}=\frac{V_{j_1, j_2}}{V}, \ldots, S_{1,2, \ldots, p}=\frac{V_{1,2, \ldots, p}}{V},$$
where $S_j$, $S_{j_1,j_2}$, for $j_1 \ne j_2$, are denoted as the first and second order sensitivity indices for input $x_j$, and interactive factors $\bx_{j_1,j_2}$ respectively. Each index measures the fractional contribution of input $x_j$ and interactions due to inputs $x_{j_1}$ and $x_{j_2}$ to the total output variance, respectively. We also define the total sensitivity index $S_j^{\top}$, an additional related measure given by $$S_j^{\top} = 1- \frac{V_{\sim j}}{V}, \quad j=1,\dots,p,$$
where $V_{\sim j} = \operatorname{var}(E\left(Y \mid \mathbf{x}_{\sim j}\right))$ is the
total variance of $\operatorname{var}\left(f_{j}\left(\mathbf{x}\right)\right)$ due to all inputs except $x_{j}$.

Recall that the posterior predictive density for the multivariate outputs from the computationally demanding model is approximated by the posterior predictive distribution using the relation $\bY^* = f(\bx^*)$ then
$$p(\bY^*|\bD) = \int \int p(\bY^*|\bTheta) \times p(\bTheta|\bD) d\bTheta,$$
where $\bD= \{\bY,\bx\}$ is the training data and $\bTheta=(\bB,\bSigma,\btau)$ are the model parameters. We also define the following identities:
\begin{equation}
E\left(Y| x_{j}\right)=\int_{\left\{x_{j'}:\{j'\neq j \}\right.} f\left(x_{1}, \ldots, x_{j}, \ldots, x_{p}\right) \prod_{\{j':\{j'\neq j\}} d G_{j'}\left(x_{j'}\right),
\end{equation}
for each specified value $x_j$ and
\begin{equation}
E(Y)=\int_{\bx} f(\bx) \prod_{j=1}^{p} dG_{j}\left(x_{j}\right)
\end{equation}
\begin{equation}
\operatorname{var}(Y)=\int_{\bx} f^{2}(\bx) \prod_{j=1}^{p} dG_{j}\left(x_{j}\right)-(E(Y))^{2}.
\end{equation}
It is possible to numerically calculate all the integrals associated with the following expectations $(E(Y))^{2}, E\left(\left(E\left(Y | x_{j}\right)\right)^{2}\right)$ and $E\left(\left(E\left(Y| \bx_{\sim j}\right)\right)^{2}\right)$. This can be obtained from the posterior distribution of MSGP  (see Appendix \ref{appC} for full details). For each MCMC iteration of the MSGP, the posterior distributions for the first-order sensitivity index, $S_{j}$, and the total sensitivity index, $S_{j}^{T},$ are computed using the equations below
\begin{multline}\label{mu0}
S_{j}= \frac{\operatorname{var}\left(E\left(Y | x_{j}\right)\right)}{\operatorname{var}(Y)}=\frac{E\left(\left(E\left(Y |x_{j}\right)\right)^{2}\right)-(E(Y))^{2}}{\operatorname{var}(Y)} \\
S_{j}^{\top}= \frac{\operatorname{var}(Y)-\operatorname{var}\left(E\left(Y | \boldsymbol{\bx}_{\sim j}\right)\right)}{\operatorname{var}(Y)} = 1-\frac{E\left(\left(E\left(Y| \bx_{\sim j}\right)\right)^{2}\right)-(E(Y))^{2}}{\operatorname{var}(Y)}.
\end{multline}

\subsection{Point estimates of the main effects}
We describe briefly the Bayesian point estimates for the main effects using the approach of \citet{q5,farah2014bayesianb}. To conserve space, let $\bY=\left(Y^{(1)}, \ldots, Y^{(m)}\right)$ and $\mathbf{f}=\left(f^{(1)}, \ldots, f^{(m)}\right)$. Given any input $\{\boldsymbol{x}=$
$\left.\left(x_{1}, \ldots, x_{p}\right), \hat Y=f(\bx)\right\}$, and as a follow up from eq. \eqref{md3}, we can determine the main effect of input $x_{j}$, but we need to calculate the following expectations $E\{E(Y) \mid D\}$ and $E\left\{E\left(Y \mid x_{ j}\right) \mid D\right\}$. This can be obtained by
$E\{E(Y) \mid D\}=\int_{{\bff}(\bx)} E(Y) p(\mathbf{f}(x) \mid D) d \bff(\bx)$
and for each specified value $x'_j$ of $j^{th}$ input
$$E\left(Y \mid x_{j'}\right)=\int_{\left\{x_{j}: j' \neq j\right\}} {f}\left(x_{1}, \ldots, x_{j'}, \ldots, x_{p}\right) \prod_{\{j: j' \neq j\}} d G_{j'}\left(x_{j'}\right)$$ and similarly,
\begin{align*}
E\{\left.E\left(Y \mid x_{j'}\right) \mid D\right\} 
=& \int_{f\left(x_{1}, \ldots, x_{j'}, \ldots, x_{p}\right)} E\left(Y \mid x_{j'}\right) p\left({f}\left(x_{1}, \ldots, x_{j'}, \ldots, x_{p}\right) \mid D\right) \\
& \times df\left(x_{1}, \ldots, x_{j'}, \ldots, x_{p}\right).
\end{align*}
See further details of implementing this procedure in \citet{farah2014bayesianb}.

\subsection{Simulation approach for computing sensitivity indices}
To compute the posterior distributions of sensitivity indices $S_{j}$ and $S_{j}^{\top}$, we shall sample from the posterior distribution of the MSGP emulator parameters. 
Using the procedure given in \citet{saltelli2002making}, generate input sample matrix $\bA_0$ of size $\bs\times p$ such that each row of $\bA_0$ is drawn independently from the input Uniform(-1,1) distribution of the simulator, $G(x) = \prod\limits_{j=1}^{p} G_{j}\left(x_{j}\right)$. In our example, we have fixed $\bs=5,000$ and recall that $p=24$.
We also generate $p$ input sample matrices, $\bA_{j}$, for $j=1, \ldots, p$ of size $\bs \times p$ each, such that the $j^{th}$ column of matrix $\bA_{j}$ equals the $j^{th}$ column of matrix $\bA_0$.
Similarly, another $p$ input sample matrices, $\bA_{\sim j}$, for $j=1, \ldots, p$ of size $\bs \times p$ each, where matrices $\bA_j$ and $\bA_0$ have all columns in common except at the $j^{th}$ column can also be generated.

Using the MSGP emulator, and for each MCMC posterior sample $\bTheta=\left\{\bB, \bSigma, \boldsymbol{\btau}\right\}$, compute the following posterior predictive samples and for each row $s$ of $\bA_0$, calculate both $\hat Y_{s}$ and $\hat Y_{s}^{2}$. 

Then for each row $s$ of $\bA_{j}$ and $s$ of $\bA_{\sim j}$, we can sample both $\hat Y_{s, j}^{\prime}$ and $\hat{Y}_{s,\sim j}^{\prime}$, respectively. For $j=1, \ldots, p$, estimate the following expectations using the posterior samples: 
\begin{align*}
&E(Y) \approx \bs^{-1} \sum\limits_{s=1}^{\bs} \hat{Y}_{s}, \quad   &E\left(\left(E\left(Y \mid x_{j}\right)\right)^{2}\right) \approx \bs^{-1} \sum\limits_{s=1}^{\bs} \hat{Y}_{s} \hat{Y}_{s, j}^{\prime},\\
&E\left(Y^{2}\right) \approx \bs^{-1} \sum\limits_{s=1}^{\bs} \hat{Y}_{s}^{2}, \quad &E\left(\left(E\left(Y \mid x_{\sim j}\right)\right)^{2}\right) \approx \bs^{-1} \sum\limits_{s=1}^{\bs} \hat{Y}_{s} \hat{Y}_{s,\sim j}^{\prime}.    
\end{align*}

By using the posterior samples above for the required expectations, we can calculate the posterior distributions for the first-order sensitivity indices, $S_{j}$ and the total sensitivity indices, $S_{j}^{\top}$ using eq. \eqref{mu0}. This procedure allows estimation of the entire distribution for each sensitivity index thereby enabling us to quantify the uncertainty of sensitivity indices.

\subsection{Extension to multivariate outputs}\label{sec1a}
In the previous section, we have assumed a univariate output and that input variables are independent which ensures the orthogonality property. Suppose that an input-output system structure is given
\begin{equation}\label{mu3a}
\bY^{(k)}=f^k = f_{0}^{(k)}+f_{j}^{(k)}+f_{\sim j}^{(k)}+f_{j, \sim j}^{(k)}
\end{equation}
and $\bY = \{Y^1,\ldots,Y^m\}$ is a m-dimensional model output vector. The Sobol decomposition for the $k^{th}$ scalar output is $\bY^{(k)}=f^{(k)}\left(x_{1}, x_{2}, \ldots, x_{p}\right).$
\citet{gamboa2013sensitivity} and \citet{lamboni2011multivariate} generalise the Sobol sensitivity indices to the case of multivariate outputs which is analogous to the univariate case by decomposing the covariance matrix $\bOmega$ of $\bY$ into the sum of partial covariance matrices $\bOmega_j$ , $\bOmega_{j,\sim j}$ and $\bOmega_{\sim j}$ respectively. This corresponds to the main contribution of $\bX_j,$ the interactive contribution of $\bx_j$ and $\bx_{\sim j}$, and the main contribution of $\bx_{\sim j}$.
Note: $ \bx_{\sim j}= \left\{x_{1}, \ldots, x_{j-1}, x_{j+1} \ldots, x_{p}\right\}$.
By taking the covariance of both sides of eq. \eqref{mu3a}, the total unconditional covariance of $\bY^{(k)}$ is given as
\begin{multline}
\operatorname{Cov}(\bY^{(k)}) = \operatorname{Cov}\left(f_{j}^{(k)}, f_{j}^{(k)}+f_{j, \sim j}^{(k)}+f_{\sim j}^{(k)}\right)+\\
\operatorname{Cov}\left(f_{j, \sim j}^{(k)}, f_{j}^{(k)}+f_{j, \sim j}^{(k)}+ f_{\sim j}^{(k)}\right)+\operatorname{Cov}\left(f_{\sim j}^{(k)}, f_{j}^{(k)}+f_{j, \sim j}^{(k)}+f_{\sim j}^{(k)}\right),
\end{multline}
where ($k=1,\ldots, m; j=1, \ldots, p$). Expanding the above expression gives

\begin{align*}
\bOmega &= \operatorname{Cov}(\bY)=\operatorname{Cov}\left(f^{(1)}, \ldots, f^{(m)}\right) \\
&=\operatorname{Cov}_{j}\left(f_{j}^{(1)}, \ldots, f_{j}^{(m)}, \bff\right)+  \operatorname{Cov}_{j, \sim j}\left(f_{j, \sim j}^{(1)}, \ldots, f_{j, \sim j}^{(m)}, \bff\right) \\
&+\operatorname{Cov}_{\sim j}\left(f_{\sim j}^{(1)}, \ldots, f_{\sim j}^{(m)}, \bff\right) \\
&= \bOmega_j + \bOmega_{j,\sim j} + \bOmega_{\sim j}
\end{align*}
and
$\bff = \left(f_{j}^{(1)}, \ldots, f_{j}^{(m)}\right) + \left(f_{j \sim j}^{(1)}, \ldots, f_{j, \sim j}^{(m)}\right) + \left(f_{\sim j}^{(1)}, \ldots, f_{\sim j}^{(m)}\right).$
Taking the trace we can calculate the total sensitivity indices

\begin{equation}\label{mu3}
S_{j}=\frac{\operatorname{tr}\left(\bOmega_{j} \right)}{\operatorname{tr}(\bOmega)}, \qquad
S_{j}^{\top}=\frac{\operatorname{tr}\left(\bOmega_{j}\right)+\operatorname{tr}\left(\bOmega_{j, \sim j}\right)}{\operatorname{tr}(\bOmega)},
\end{equation}
where $\operatorname{tr}(\bOmega)=\sum \limits_{k=1}^{m} \bV^{(k)}$ and $\operatorname{tr}\left(\bOmega_{j}\right)=\sum \limits_{k=1}^{m} \operatorname{Cov}\left(f_{j}^{(k)}, f_{j}^{(i)}\right)=\sum \limits_{k=1}^{m} V_{j}^{(k)} $. Therefore, $S_{j}=\frac{\sum \limits_{k=1}^{m} V_{j}^{(k)}}{\sum \limits_{k=1}^{m} \bV^{(k)}}.$
See \citet{xu2019sensitivity,xiao2016sensitivity} for further details.

\subsection{Generalised sensitivity indices for multivariate output using vector projection}
We note that the generalised Sobol sensitivity indices based on a decomposition of the covariance matrix of multiple outputs defined by \citet{gamboa2013sensitivity} and \citet{lamboni2011multivariate} above in eq. \eqref{mu3} does not take into consideration the correlation between multiple outputs. This makes it difficult to interpret the sensitivity results. To address this problem, \citet{xu2019generalized} developed new generalised indices based on the vector projection of multivariate outputs for measuring the comprehensive effects of the inputs.
The method is based on vector projection and dimension normalisation. The approach involves an affine coordinate transformation of the normalised outputs based on the correlations among different variables. Firstly, the multivariate outputs are normalised into a dimensionless unit to remove the interference of different dimensions. Secondly, an affine coordinate system is computed by using the correlations among the different outputs. To obtain new sensitivity indices, we can project the variance contribution vector composed by the individual variance contribution of the input to each output on the total variance vector. The new sensitivity indices are defined as the ratio of the projection of the variance contribution vector to the norm of the total variance vector.

To obtain dimensionless outputs, the output $\bY^{k}$ can be normalized as
$\bY^{'k}=\frac{\left(\bY-\bar \bY\right)} {\sigma(\bY)}$, where $\sigma$ is the standard deviation. Let
\begin{equation*}
R(\bY') =
\begin{pmatrix}
r_{1,1} & r_{1,2} & \cdots & r_{1,m} \\
r_{2,1} & r_{2,2} & \cdots & r_{2,m} \\
\vdots & \vdots & \ddots & \vdots \\
r_{m,1} & r_{m,2} & \cdots & r_{m,m}
\end{pmatrix}
\end{equation*}
be the correlation coefficient matrix of normalized multivariate output data $\bY'$, where $r_{i,k}= \frac{\operatorname{Cov}(\bY^i,\bY^k)}{\sqrt(var(\bY^i)\times var(\bY^k))}, (i, k: 1,\ldots,m)$. Let $\theta^{(i, k)} = \mbox{arccos}(r_{(i,k)})$ be the angle expanded by $\bY^{(i)}$ and $\bY^{(k)}$. Then, $m-$ dimensional outputs $\bY^{(i)}(i = 1,\ldots, m)$ expand an affine coordinate system, respectively, with the unit directional vector $\be_i(i=1, \ldots, m)$, and $\be_i^T\be_k = \operatorname{cos}(\theta^{(i),(k)}= r_{i,k}(i, k=1,\ldots, m)$. From the definition of the angle $\theta^{(i),(k)}$ expanded by $\bY^{(i)}$ and $\bY^{(k)}$ as $\theta^{(i, k)} = \operatorname{arccos}(r_{(i,k)}$ the $m-$ dimensional affine coordinate system can be determined competely in which the correlations of the multivariate outputs are incorporated. Having constructed an affine coordinate system, the individual variances of each output can be viewed as vectors $\bOmega_j^{(\bY^{i})} \be_i(i=1, \ldots, m)$, and the total variance vector of $m-$ dimensional outputs can be obtained as $\bOmega_j = \bOmega^{\bY^{(1)}}\be_1+ \bOmega^{\bY^{(2)}}\be_2+ \ldots + \bOmega^{\bY^{(m)}}\be_m$, for $(j=1,\ldots,p)$

We define the matrix $\bQ_j$ as
\begin{equation*}
\bQ_j= \Vert\bOmega_j\Vert \cos \theta_j = \frac{\langle \bOmega_{j},\bOmega\rangle}{\Vert\bOmega\Vert}
=\frac{ \bigg[\bOmega_j^{(1)},\ldots \bOmega_j^{(m)}\bigg] R(\bY')\bigg[\bOmega^{(1)},\ldots \bOmega^{(m)}\bigg]^T}{\sqrt\bigg[ \bOmega^{(1)},\ldots \bOmega^{(m)}\bigg] R(\bY')\bigg[\bOmega^{(1)},\ldots \bOmega^{(m)} \bigg]^T},
\end{equation*}
and $P_j$ is the projected first order variance contribution which is analogous to the $S_{j}$ defined in eq. \eqref{mu3} and corresponding expression for the total indices $S_{j}^{T}$ is given in Appendix \ref{appC}.
\begin{equation}\label{mu4}
\bP_j= \frac{\bQ_j}{\Vert \bOmega \Vert} = \frac{\langle \bOmega_{j},\bOmega\rangle}{\Vert\bOmega\Vert^2},
\end{equation}
where $\theta_j$ is the angle between between $\bOmega$ and $\bOmega_j$, $\langle.,.\rangle$ denotes the inner product of two vectors and $\Vert.\Vert$ is the norm of a vector.
The notable properties of the vector projection based sensitivity indices are given below:
\begin{itemize}
\item $\left|\theta_{j}\right| \leq \frac{\pi}{2}, 0 \leq \hat{\bP}_{j} \leq 1 .$
\item Suppose there is an input $X_{i}$ which has no effect on $\bY$ but an input $X_{j}$ has an effect on $\bY$, then $\hat{\bP}_{i, j}=\hat{\bP}_{j}$.
\item Similarly, if $X_{i}$ has no effect on $\bY$, then the total projected indices $\hat{\bP}^T_{i}=0$.
\item $\frac{\bOmega_j}{|\bOmega|}=\frac{\bOmega^{\bY(1)}}{\bOmega}\be_1 + \frac{\bOmega^{\bY(2)}}{\bOmega}\be_2 + \ldots \frac{\bOmega^{\bY(m)}}{\bOmega}\be_m.$
\end{itemize}
See further details in \citet{xu2019sensitivity} and \citet{xu2019generalized}.

It is worth noting that variance-based decomposition is not the only approach for performing sensitivity indices.
Alternative approaches exist for identifying essential variables using Gaussian processes. For multivariate Gaussian process emulators, input variables can impact the response through both their mean functions and the correlation structure. Hence, in our model description above, the relative importance of the input variables can also be determined by the relative magnitude of the corresponding correlation parameters $\btau$. A large value for $\btau_k$ indicates a weak correlation. This approach was used by \citet{liu2009dynamic} and is very similar to automatic relevance
determination. In this approach, the inverse length scale parameter of each input variable could be used as a proxy for variable relevance \citet{williams1996gaussian,paananen2019variable}. This method is also called sparse Bayesian learning in some literature because of its capability to learn sparse solutions to linear problems, or it can also be viewed as an in-built mechanism for feature selection.
Another related approach to automatic relevance determination for active variable selection is through the sparsifying spike-and-slab priors \citet{savitsky2011variable} and in the work of \citet{crawford2019variable} where Kullback-Leibler divergence between the marginal distribution of the rest of the variables to their conditional distribution is fixed to a nominal distribution.

\section{Results}
\subsection{Test cases}
\begin{figure}[!htb]
\begin{center}
\includegraphics[width=.99\textwidth]{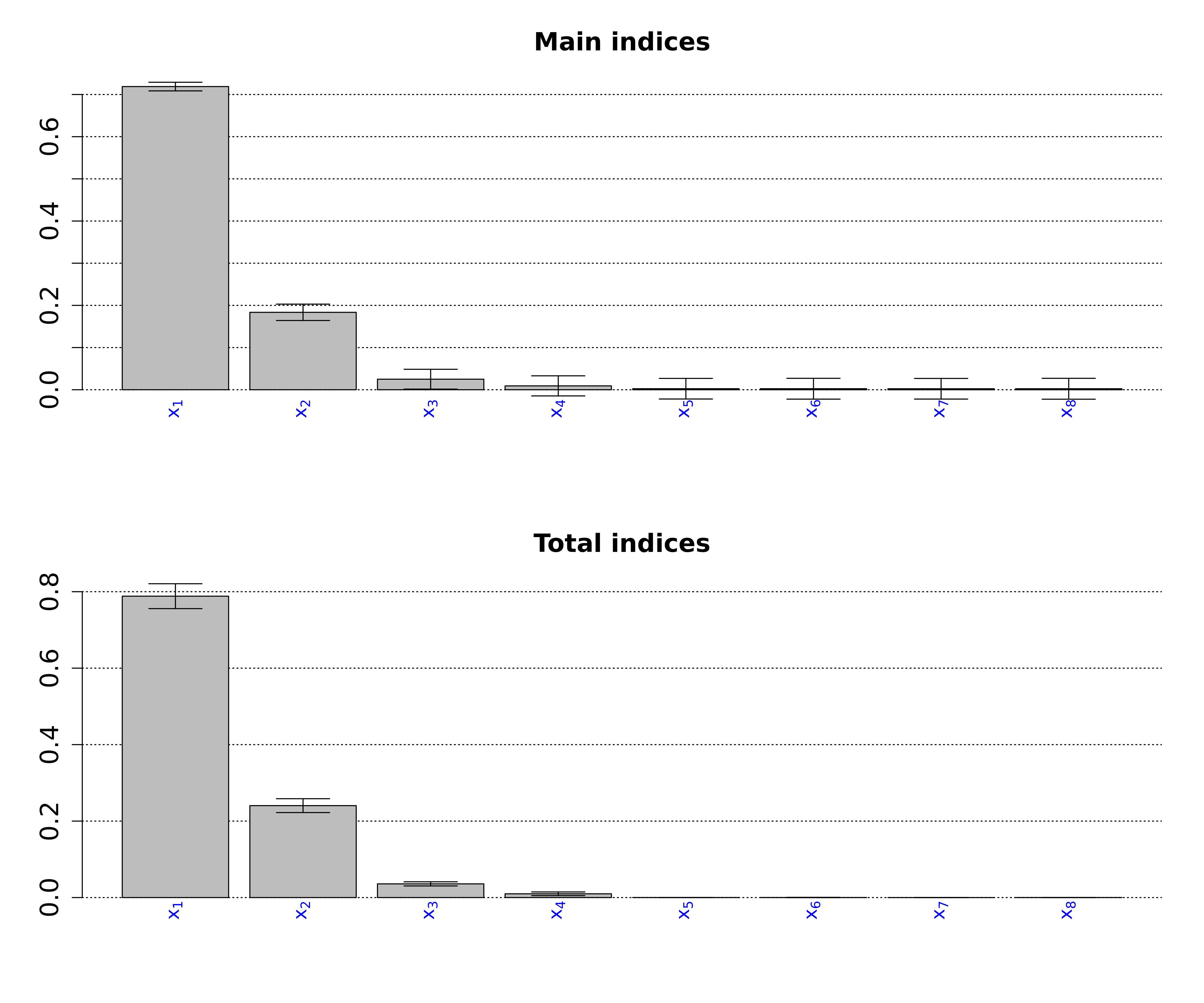}
\end{center}
\caption{Main and total indices for the Sobol g function using our algorithm.}\label{figtest}
\end{figure}

We illustrate our sensitivity algorithm with two test cases:\\
(i) An analytic function called the Sobol $g-$function (non-correlated input) is a common example used in global sensitivity analysis \citep{saltelli2002making}. The $g$ function is a strongly non-monotonic, non-additive function of independent factors $x_{i},$ assumed identically and uniformly distributed in the unit cube $I^{p}=\left\{x \mid 0 \leqslant x_{i} \leqslant 1 ; i=1,2, \ldots, p\right\}$
$$g\left(x_{1}, x_{2}, \ldots, x_{p}\right)=\prod_{i=1}^{p} g_{i}\left(x_{i}\right),$$
where $g_{i}\left(x_{i}\right)=\frac{\left|4 x_{i}-2\right|+a_{i}}{1+a_{i}}$.
The Sobol $g-$function was also used in the works of \citet{gamboa2020global} to perform the numerical comparisons between the standard Pick-Freeze approach and Chatterjee's estimators. In this case, we consider $p=8$ factors, with $a = (0, 1, 4.5, 9, 99, 99, 99, 99)$.
We can see from Figure~\ref{figtest}(a) that the lower the coefficient $a$, the larger the sensitivity indices. This result is comparable to the approach implemented in the R package \textit{sensitivity} using the function \textit{sobolshap-knn} which computes both the Sobol indices and Shapley effects \citep{broto2020variance}.\\

\begin{figure}[!htb]
\begin{minipage}{.5\textwidth}
\centering
\includegraphics[width=.99\linewidth]{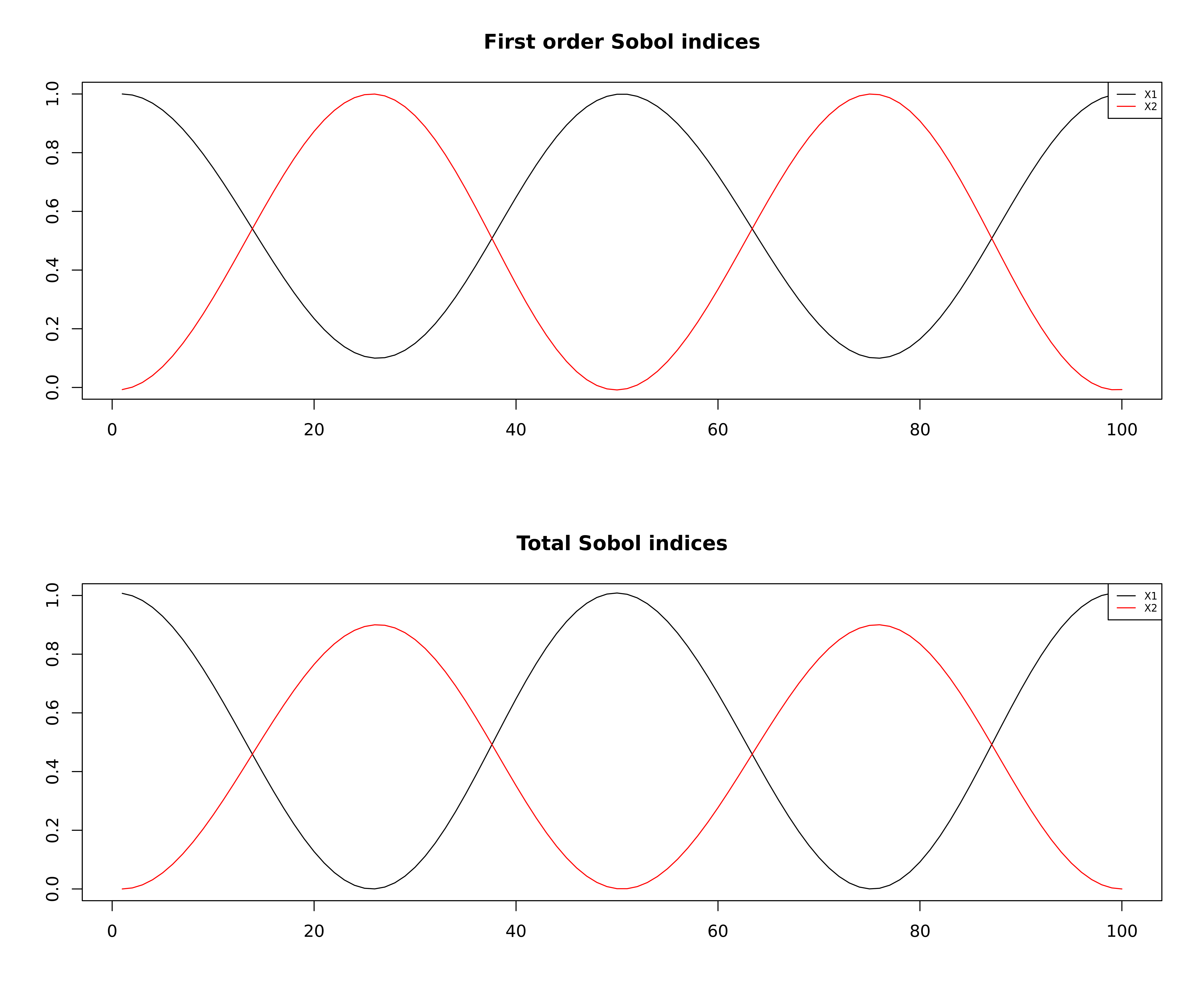}
Standard results (a)
\end{minipage}
\begin{minipage}{0.5\textwidth}
\centering
\includegraphics[width=.99\linewidth]{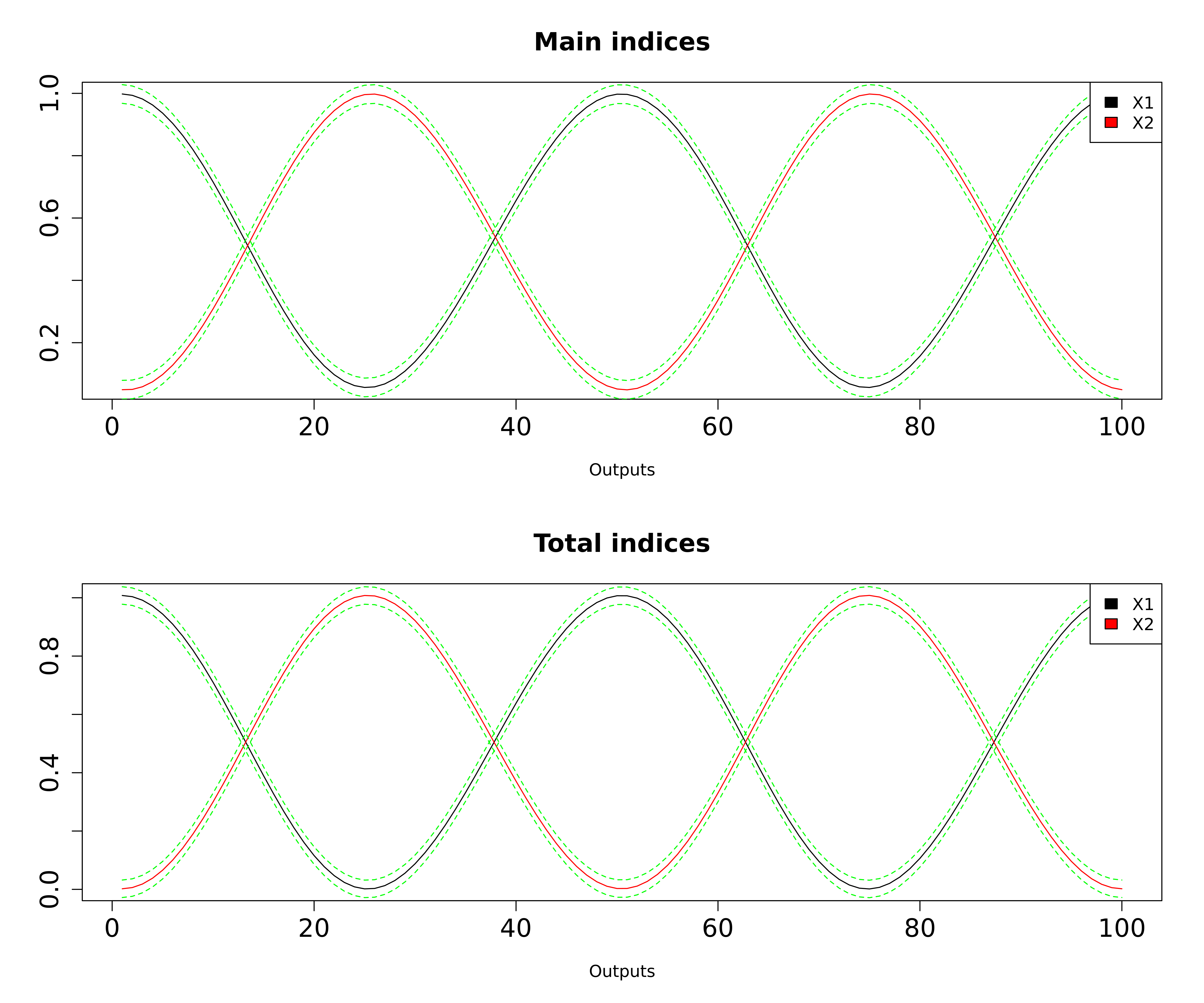}
MSGP results (b)
\end{minipage}
\caption{Main and total indices for the multivariate test cases using arctangent temporal function which has 100 functional outputs with two input variables from a uniform distribution.}\label{figtest2}
\end{figure}

(ii) The second example is a multivariate case study using the arctangent temporal function with two independent variables and 100 outputs. The functional support is on $[0,2\pi]$, where $q$ is the number of discretization steps of the functional outputs \citep{auder2011classification}. There are two factors, all following the uniform distribution between $[-7,7]$. The result is given in Figure~\ref{figtest2} below showing the comparison between the standard approach and MSGP for both main and total indices. We could see that both results are similar.


\subsection{Application to IAP2 simulation}
\begin{figure}[!htb]
\begin{minipage}{.5\textwidth}
\centering
\includegraphics[width=.9\linewidth]{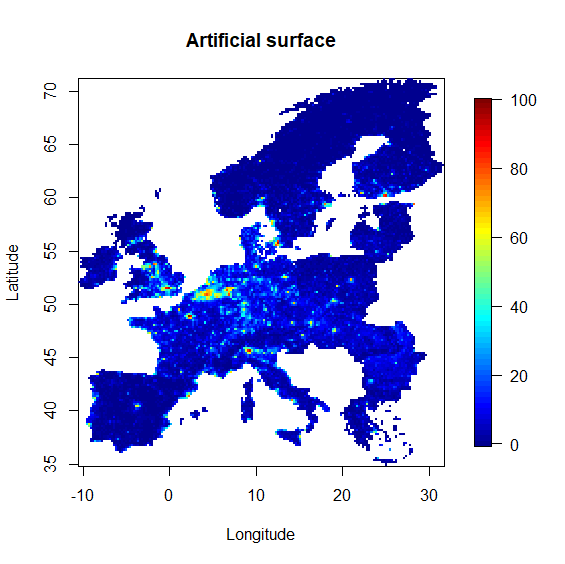}
Artificial surface for SSP3 scenario in 2050s (a)
\end{minipage}
\begin{minipage}{0.5\textwidth}
\centering
\includegraphics[width=.9\linewidth]{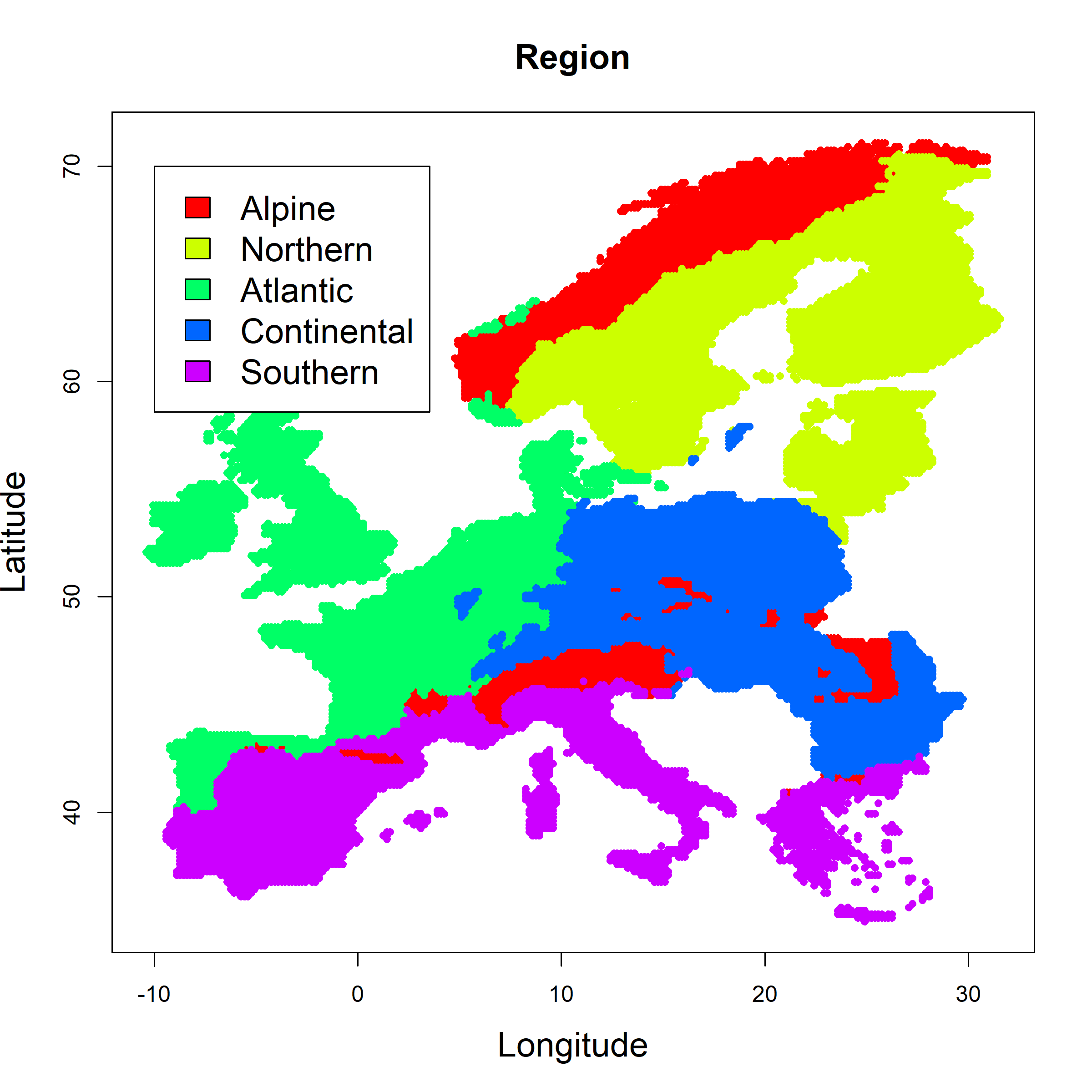}
Regional classification (b)
\end{minipage}
\caption{Sample output data from the IMPRESSIONS IAP2 simulation under RCP4.5 and HadGEM2-ES-RCA4 climate model and the classification of the five European regions used in the analyses.}\label{fig1}
\end{figure}

We first describe the procedure for building the MSGP then apply the MSGP method to compute sensitivity indices of the IAP2 simulation models under various scenarios to explore a wide range of uncertainty.
The effect of model inputs was evaluated on all 18 spatially-distributed model outputs using MSGP.
We note that it is practically infeasible to perform a fully spatio-temporal sensitivity analysis for the IAP2 model.
We, therefore, follow the works of \citet{kebede2015direct} and \citet{savall2019sensitivity}, where we aggregated the output data into six spatial extents (see Figure \ref{fig1}(b)), namely Europe-wide and five European regions to reduce computational expense. We subsampled 600 data points from each of the five scenarios (baseline, SSP1, SSP3, SSP4, SSP5) and three timeslices (2020s, 2050s, 2080s) giving 9,000 ($5 \times 3 \times 600$) samples per each output. The training data $\bD=\{\bX_{n\times p}, \bY_{n\times m} \}$, where $n=9,000, p=24$ and $m=18$ for different cases was considered in this paper. We used the multivariate sparse Gaussian process (MSGP) described in Section~\ref{senn1} to fit the data and compute the new sensitivity indices using the procedure highlighted in Section~\ref{senn2}.

\subsubsection{A procedure for building the MSGP}
\begin{table}
\caption{\small Model summary: crossvalidated proportion of variance P, root mean squared error $\rho$ showing the performance of the MSGP emulators for Matern$_{5/2}$ with Wendland correlation functions; and with Truncated power correlation for four different levels of sparsity for the EU-wide average data. \label{sparse0}}
\centering
\scalebox{0.8}{
\begin{tabular}{|l|rr|rr|rr|rr|rr|}
\hline
Models &1&&2& &3& &4& &5&\\
 \hline
$\omega$ ($\%$) &\multicolumn{1}{p{2cm}}{\centering Matern\\ Wendland}&&80 &&90& &95& &99&\\
\hline
\hline
Output & P & $\rho$ &P &$\rho$ &P &$\rho$ &P &$\rho$ &P &$\rho$\\
\hline
Food Capita & 0.54 & 0.68 & 0.52 & 0.70 & 0.52 & 0.46 & 0.52 & 0.46 & 0.51 &0.70 \\
  People Flooded  & 0.96 & 0.21 & 0.95 & 0.23 & 0.95 & 0.17 & 0.95 & 0.17 & 0.95 & 0.23 \\
  Water Expl Ind & 0.40 & 0.78 & 0.34 & 0.81 & 0.34 & 0.59 & 0.34 & 0.59 & 0.34 & 0.81 \\
  Landuse Int Ind & 0.82 & 0.42 & 0.81 & 0.44 & 0.81 & 0.33 & 0.81 & 0.33 & 0.81 & 0.44 \\
  Landuse Diversity & 0.76 & 0.49 & 0.74 & 0.51 & 0.74 & 0.38 & 0.74 & 0.38 & 0.74 & 0.51 \\
  Biodiversity & 0.55 & 0.68 & 0.47 & 0.73 & 0.47 & 0.58 & 0.47 & 0.58 & 0.47 & 0.74 \\
  Timber Prod & 0.69 & 0.56 & 0.66 & 0.58 & 0.66 & 0.48 & 0.66 & 0.48 & 0.66 & 0.59 \\
  Artificial Surfaces & 0.47 & 0.73 & 0.33 & 0.82 & 0.33 & 0.69 & 0.33 &0.69 & 0.34 & 0.82 \\
  Food Prod & 0.73 & 0.52 & 0.66 & 0.59 & 0.66 & 0.59 & 0.66 & 0.59 & 0.66& 0.58 \\
  Carbon Stock & 0.61 & 0.63 & 0.56 & 0.66 & 0.56 & 0.66 & 0.56 & 0.66 & 0.57 & 0.66 \\
  Irrigation Usage & 0.57 & 0.65 & 0.51 & 0.70 & 0.51 & 0.70 & 0.51 & 0.70 & 0.52 & 0.68 \\
  Intensive Arable & 0.68 & 0.57 & 0.65 & 0.59 & 0.65 & 0.59 & 0.65 & 0.59 & 0.65 & 0.59 \\
  Intensive Grass & 0.74 & 0.51 & 0.69 & 0.56 & 0.69 & 0.56 & 0.69 & 0.56 &0.69 & 0.56 \\
  Extensive Grass & 0.54 & 0.68 & 0.52 & 0.70 & 0.52 & 0.70 & 0.52 & 0.70 & 0.53 & 0.69 \\
  Very Ext Grass & 0.39 & 0.78 & 0.32 & 0.83 & 0.32 & 0.83 & 0.32 & 0.83 & 0.32 & 0.82 \\
  Unmanaged Land & 0.55 & 0.67 & 0.52 & 0.69 & 0.52 & 0.69 & 0.52 & 0.69 & 0.53 & 0.70 \\
  Managed Forest & 0.65 & 0.59 & 0.61 & 0.63 & 0.61 & 0.63 & 0.61 & 0.63 & 0.61& 0.63 \\
  Unmanaged Forest & 0.77 & 0.48 & 0.74 & 0.51 & 0.74 & 0.51 & 0.74 & 0.51 & 0.75 & 0.50 \\
   \hline
\end{tabular}
}
\end{table}

We fit the MSGP by first by spatially averaging the entire output data (called EU-wide average datasets) and rescale the input data to the range [-1, 1]. The input data in this analysis represent the 24 model parameters. We scaled the data by centering the column data with their respective minimum values and divided by their range, such that $\bx'=\bigg(2\times\frac{\left(\bx-x_{\min }\right)} {\left(x_{\max }-x_{\min }\right)}\bigg) -1$. 
The output data was transformed by normalising it to have a zero mean and unit variance. Standardising these data around the centre 0 with a standard deviation of 1 is essential because the simulated outputs have different units. This is a necessary step in the data preprocessing to meet model assumptions and reduce bias in the estimated sensitivity results.
This procedure is essential to reduce the effect of magnitude/unit of the response variables on the computed sensitivity indices. We note that variables with larger magnitude would make larger contributions to the overall output variances. 

\begin{figure}[!htb]
\begin{center}
\includegraphics[width=.99\textwidth]{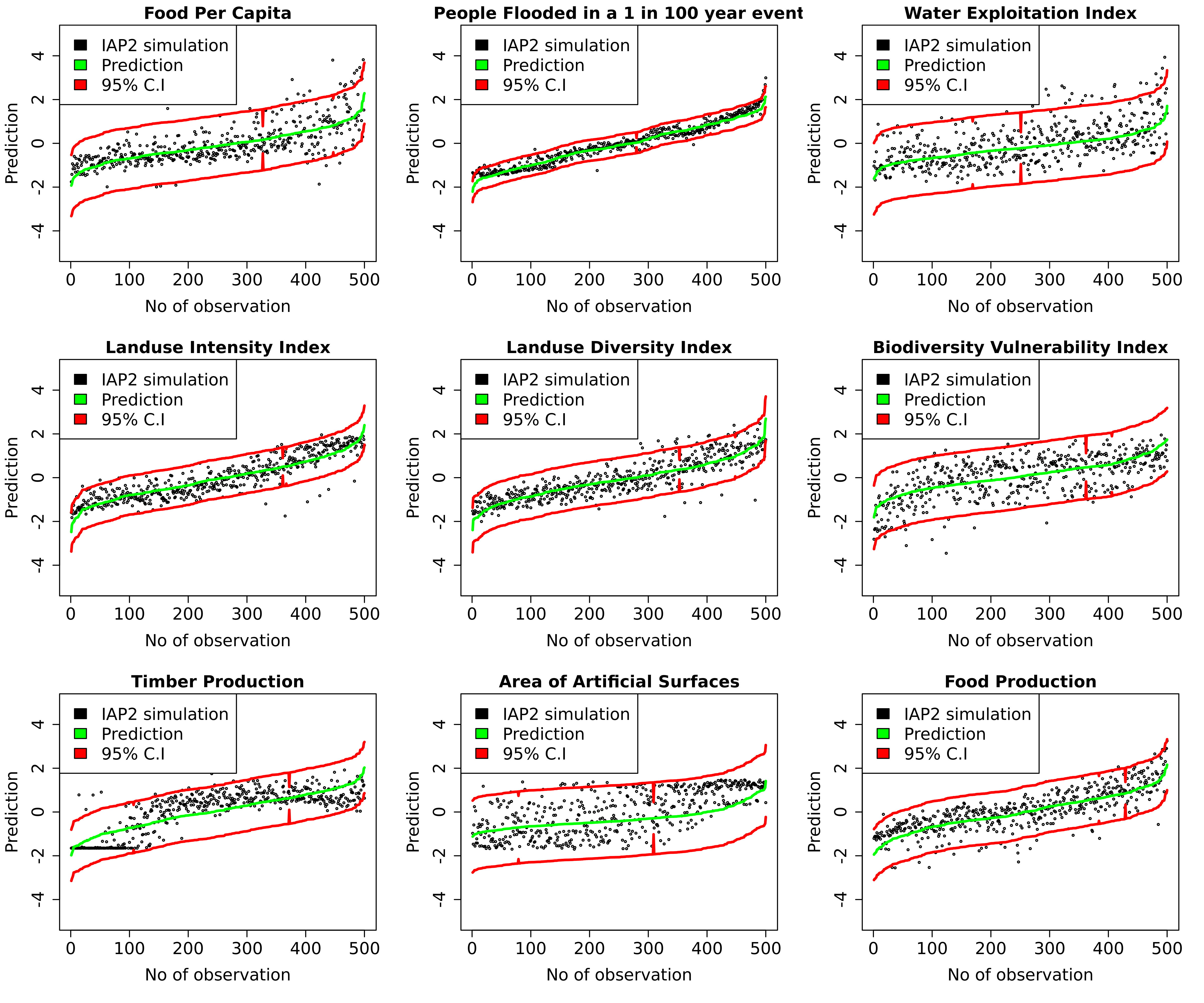}
\end{center}
\caption{Comparison of the multivariate GP emulator performance with IAP2 simulation for 500 randomly sample points under the SSP4 scenario in the 2050s for EU-wide aggregated data; simulation (black) and emulator predictions (green) with 95\% C.I. (red). Note: the outputs have been normalised to have a zero mean and unit variance.}\label{fig4a}
\end{figure}

\begin{figure}[!htb]
\begin{center}
\includegraphics[width=.99\textwidth]{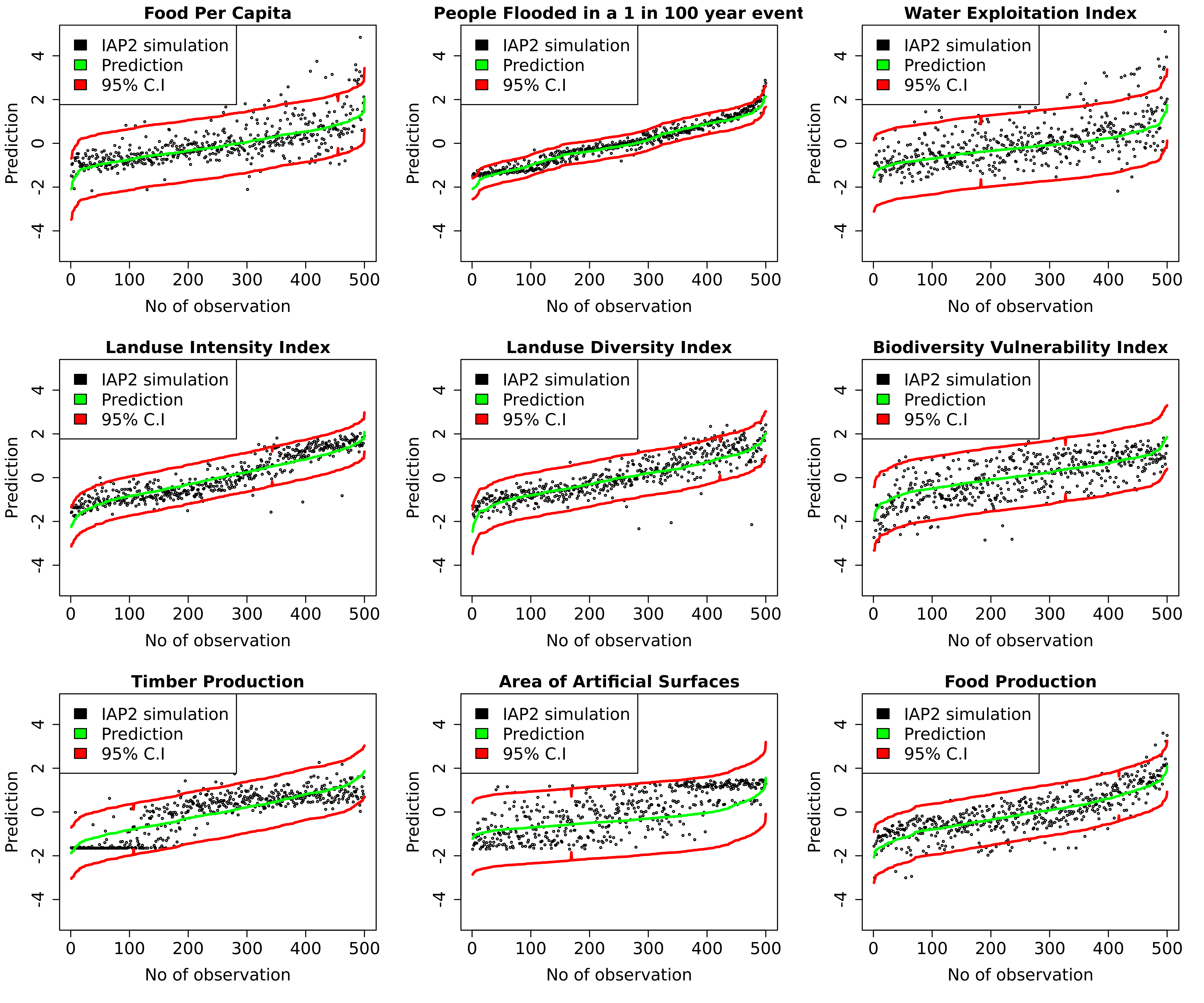}
\end{center}
\caption{Comparison of the multivariate GP emulator performance with IAP2 simulation for 500 randomly sample points under the SSP1 scenario in the 2050s for EU-wide aggregated data; simulation (black) and emulator predictions (green) with 95\% C.I. (red). Note: the outputs have been normalised to have a zero mean and unit variance.}\label{fig4b}
\end{figure}

The model parameters are estimated using an MCMC procedure (see Section \ref{sec:sampling}) to approximate the posterior distributions of the unknown parameters $\widehat \Theta=\left\{ \widehat \bB, \widehat \bSigma, \hat \btau,\right\}$ using the MNIW prior.
The parameter $\btau$ is sampled using an adaptive Metropolis-Hastings algorithm with a multivariate normal random walk proposal distribution centred around the most current estimates of $\btau$. We fixed the target acceptance rate to the recommended 0.234. See \citet{vihola2012robust} for full details. 
We assess the convergence of the MCMC procedure using standard diagnostic tests (see Appendix \ref{appE}). For example, Table~\ref{est_param} in Appendix \ref{appE} gives the potential scale reduction factor (PSRF), which is an estimated factor by which the scale of the current distribution for the target distribution can be reduced if the simulations are continued for a large number of samples. It can also be used as an estimate for the R-hat metric \citep{vehtari2019rank}. To compute these values, we ran three parallel chains with different starting values. We sampled 50,000 MCMC iterations with a thinning frequency of 25 using the algorithm described in the previous section. We discarded the first 1,000 samples as burn-in. We see that the PSRF values for most of the parameters are very close to 1, which indicates they are likely to converge to the correct target distribution. According to the literature, a recommended approach is to achieve PSRF < 1.1, although this depends on the nature of the problem. A large PSRF > 1.2 indicates no convergence and the possible presence of a multimodal marginal
posterior distribution in which different chains may have converged to different modes. See \citet{gelman1992inference,brooks1998general} for further details. We also included both the density and trace plots of the estimated parameter $\btau$ in Figures~\ref{hist_param} and \ref{trace_param} respectively in Appendix \ref{appE} as a further visual diagnostic. We note that the magnitude of the estimated scale $\btau$ parameters is also a proxy for sensitivity indices. 

To implement the above procedure on the EU-wide average data, we replaced the dense $n \times n$ matrix $\mathfrak{R}$ with a sparse approximation using compactly supported correlation functions such as the Bohman and truncated power functions and products of Matern$_{5/2}$ and Wendland functions, as described in Appendix \ref{appB}. It is infeasible to jointly estimate the levels of sparsity $\omega$ (percentage of off-diagonal nonzero elements), the cut-off $c$ and other parameters. However, it is possible to fix these at reasonable values with relatively little effects on predictions and model performance.
Here, we used $K-$fold cross-validation to determine the appropriate level of sparsity to be used for the sensitivity analyses in the next section. We used root mean square errors ($\rho$), and proportion of variance ($P$) explained, to compare the
efficiency at fixed sample sizes. Table~\ref{sparse0} compares the results. 

We observe that all the fitted models performed relatively well with the proportion of variance explained $P$ ranging from $0.34-0.96$ and $\rho$ $0.21-0.83$ for various sparsity levels. We note that the higher the $P$ values, the better the model and vice-versa for the $\rho$. The \textit{People flooded in a 1 in 100-year event} variable has the largest $P$ and smallest $\rho$ values respectively for various degrees of sparsity and is thus more predictable than the other variables. On the other hand, the \textit{very extensive grassland} variable seems less predictable compared to the others. Generally, the models made substantial savings by using sparse models, with the time taken to fit each model decreasing with the degree of sparsity.

The results are generally similar across different levels of sparsity. We can deduce that sparsity does not necessarily degrade the performance of the models. 
The results are relatively similar, except that the models under the truncated power correlation function are much faster to train. We have fixed the value of $\omega$ as $90\%$, which minimised the $\rho$ and maximised the $P$ values for subsequent analyses in this paper.

Figures~\ref{fig4a} and \ref{fig4b} are the graphical comparison between the multivariate MSGP emulator with the IAP2 simulation for different outputs for model 3 in Table~\ref{sparse0}. The results are shown for the SSP4 and SSP1 scenarios in the 2050s for EU-wide aggregated data. The emulators reproduce the patterns well in most of the variables, with predictions falling within the 95\% C.I. For instance, the variable \textit{people flooded in a 1 in 100-year event} has a narrow confidence band. The predictions are relatively precise in most of the points. As suggested by Table~\ref{sparse0}, and seen again here, this variable is well-predicted compared to other variables. The \textit{food per capita} and \textit{water exploitation index} are relatively better predicted compared to the \textit{land use diversity index} with several points fall outside the confidence bands in Figures~\ref{fig4a}. There are some mismatches between the emulator predictions and simulations in \textit{timber production}. The results are also similar in Figure~\ref{fig4b}.

We now consider the regional analysis. Five different MSGP models were also fitted, each for the five European regional aggregates. This analysis allows us to compute regional sensitivity indices. Table~\ref{ref4} shows the model comparison. The results are generally similar across the regions. The \textit{people flooded in a 1 in 100-year event, water exploitation index} and \textit{artificial surfaces} are still well-predicted compared to the other variables, with high $P$ and low $\rho$ values, respectively. The corresponding regional sensitivity indices are provided in Figure~\ref{fig8} in Appendix \ref{appA}.

\begin{table}
\caption{Model summary: crossvalidated proportion of variance P, root mean squared error $\rho$ showing the performance of MSGP emulators for 5 regional aggregations for all the scenarios and timeslices.  AL=Alpine, NO=Northern; AT=Atlantic; CO=Continental; SO=Southern.}
\label{ref4}
\centering
\scalebox{0.8}{
\begin{tabular}{|l|rr|rr|rr|rr|rr|}
\hline
& AL&&NO&&AT&&CO&&SO&\\
\hline
&   P&$\rho$&P&$\rho$&P&$\rho$ &P&$\rho$&P&$\rho$\\
  \hline
Food Capita & 0.41 & 0.78 & 0.53 & 0.69 & 0.41 & 0.77 & 0.61 & 0.62 & 0.79 & 0.46 \\
  People Flooded & 0.76 & 0.49 & 0.79 & 0.46 & 0.88 & 0.35 & 0.75 & 0.50 & 0.77 & 0.48 \\
  Water Exploit & 0.35 & 0.81 & 0.25 & 0.87 & 0.29 & 0.84 & 0.35 & 0.80 & 0.32 & 0.83 \\
  Landuse Inten & 0.70 & 0.55 & 0.70 & 0.55 & 0.83 & 0.41 & 0.78 & 0.47 & 0.56 & 0.67 \\
  Landuse Diver & 0.64 & 0.60 & 0.69 & 0.56 & 0.30 & 0.84 & 0.57 & 0.66 & 0.58 & 0.65 \\
  Biodiversity Vul & 0.54 & 0.68 & 0.40 & 0.78 & 0.46 & 0.74 & 0.47 & 0.73 & 0.61 & 0.63 \\
  Timber Prod & 0.54 & 0.68 & 0.55 & 0.67 & 0.55 & 0.67 & 0.72 & 0.54 & 0.41 & 0.76 \\
  Artificial Surfaces & 0.32 & 0.82 & 0.32 & 0.83 & 0.32 & 0.82 & 0.35 & 0.81 & 0.33 & 0.82 \\
  Food Prod & 0.49 & 0.72 & 0.57 & 0.66 & 0.49 & 0.71 & 0.70 & 0.54 & 0.72 & 0.53 \\
  Carbon Stock & 0.34 & 0.82 & 0.49 & 0.72 & 0.51 & 0.71 & 0.57 & 0.66 & 0.29 & 0.85 \\
  Irrigation  & 0.41 & 0.77 & 0.27 & 0.86 & 0.49 & 0.72 & 0.50 & 0.71 & 0.45 & 0.74 \\
  Intensive Arable & 0.46 & 0.73 & 0.63 & 0.61 & 0.40 & 0.77 & 0.59 & 0.64 & 0.45 & 0.74 \\
  Intensive Grass & 0.60 & 0.64 & 0.42 & 0.77 & 0.76 & 0.50 & 0.54 & 0.68 & 0.70 & 0.55 \\
  Extensive Grass & 0.44 & 0.75 & 0.32 & 0.83 & 0.44 & 0.74 & 0.51 & 0.70 & 0.38 & 0.79 \\
  Very Ext. Grass & 0.30 & 0.84 & 0.24 & 0.87 & 0.34 & 0.82 & 0.20 & 0.90 & 0.40 & 0.77 \\
  Unmanaged Land & 0.62 & 0.62 & 0.57 & 0.66 & 0.44 & 0.74 & 0.37 & 0.79 & 0.30 & 0.83 \\
  Managed For & 0.54 & 0.68 & 0.55 & 0.67 & 0.53 & 0.68 & 0.69 & 0.56 & 0.42 & 0.76 \\
  Unmanaged For & 0.67 & 0.58 & 0.61 & 0.63 & 0.71 & 0.54 & 0.71 & 0.54 & 0.61 & 0.63 \\
   \hline
\end{tabular}
}
\end{table}

\subsection{Sensitivity results}
For the practical issue of computationally demanding computer models, we show that the substitution of the original model by the MGSP models facilitates the rapid estimation of these global sensitivity indices with precision at a reasonable computational cost.

\begin{figure}[!htb]
\begin{center}
\includegraphics[width=.99\textwidth]{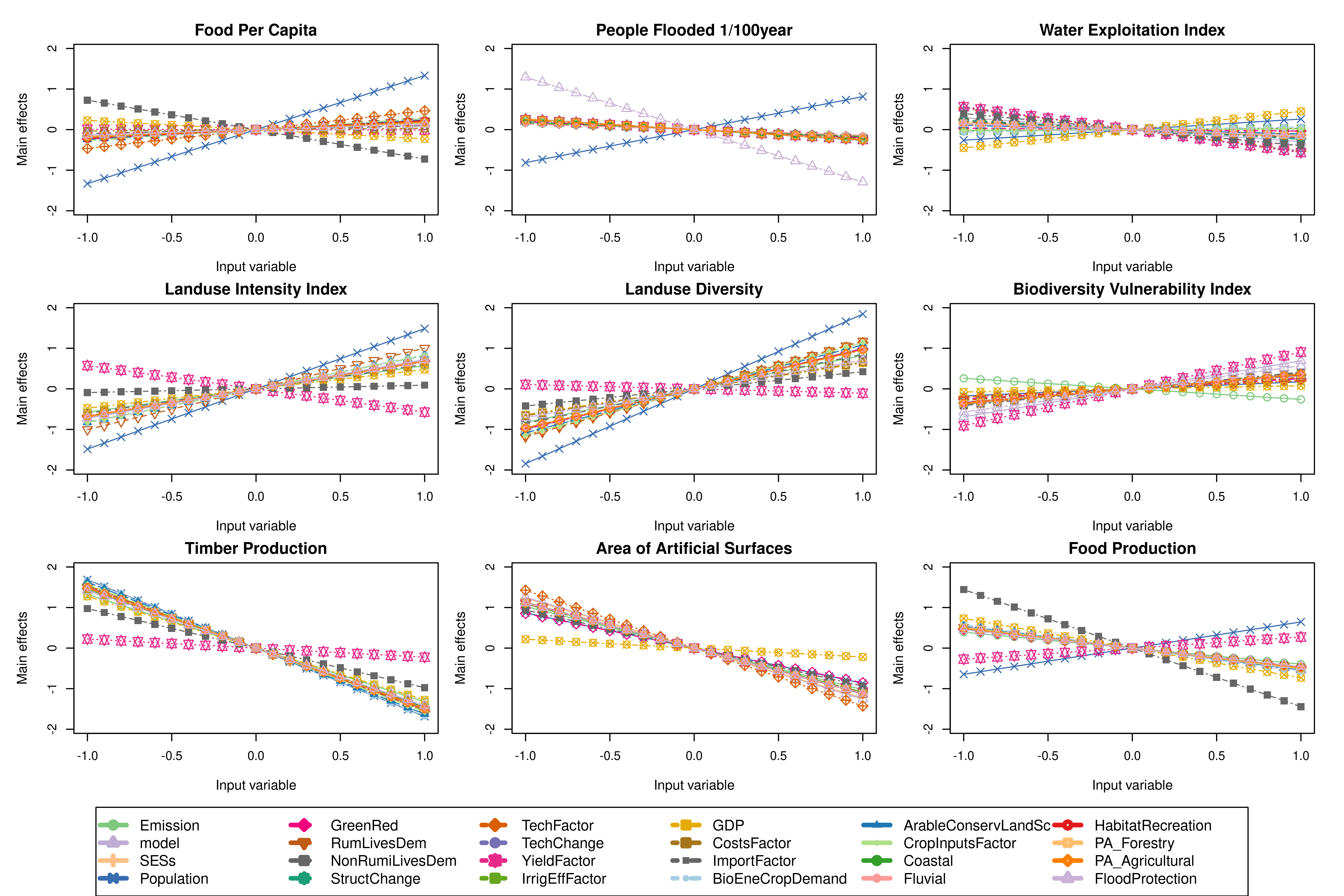}
\end{center}
\caption{Main effect plots for EU-wide aggregated data; simulation (black) and emulator predictions (green) with 95\% C.I. (red). Note: the outputs have been normalised to have a zero mean and unit variance.}\label{fig4aa}
\end{figure}

Figure~\ref{fig4aa} shows the main effects for all the input variables for some of the outputs. The input value is normalized between $[-1,1]$. A large variation in the main effect plot is an indication of the greater influence of such input on the output. We see that population has a greater influence on \textit{food per capita, people flooded in 1 in 100-year landuse intensity} and \textit{landuse diversity} indices. An increase in population causes these indices to increase. Import factor has a negative effect on \textit{food per capita, timber and food production} while the influence of yield factor also reduces with an increase in \textit{landuse intensity index}. The plots for the remaining outputs are given in Figure~\ref{fig4aa2} of Appendix \ref{appA}.

\begin{figure}[hbt!]
\centering
\includegraphics[width=1\linewidth]{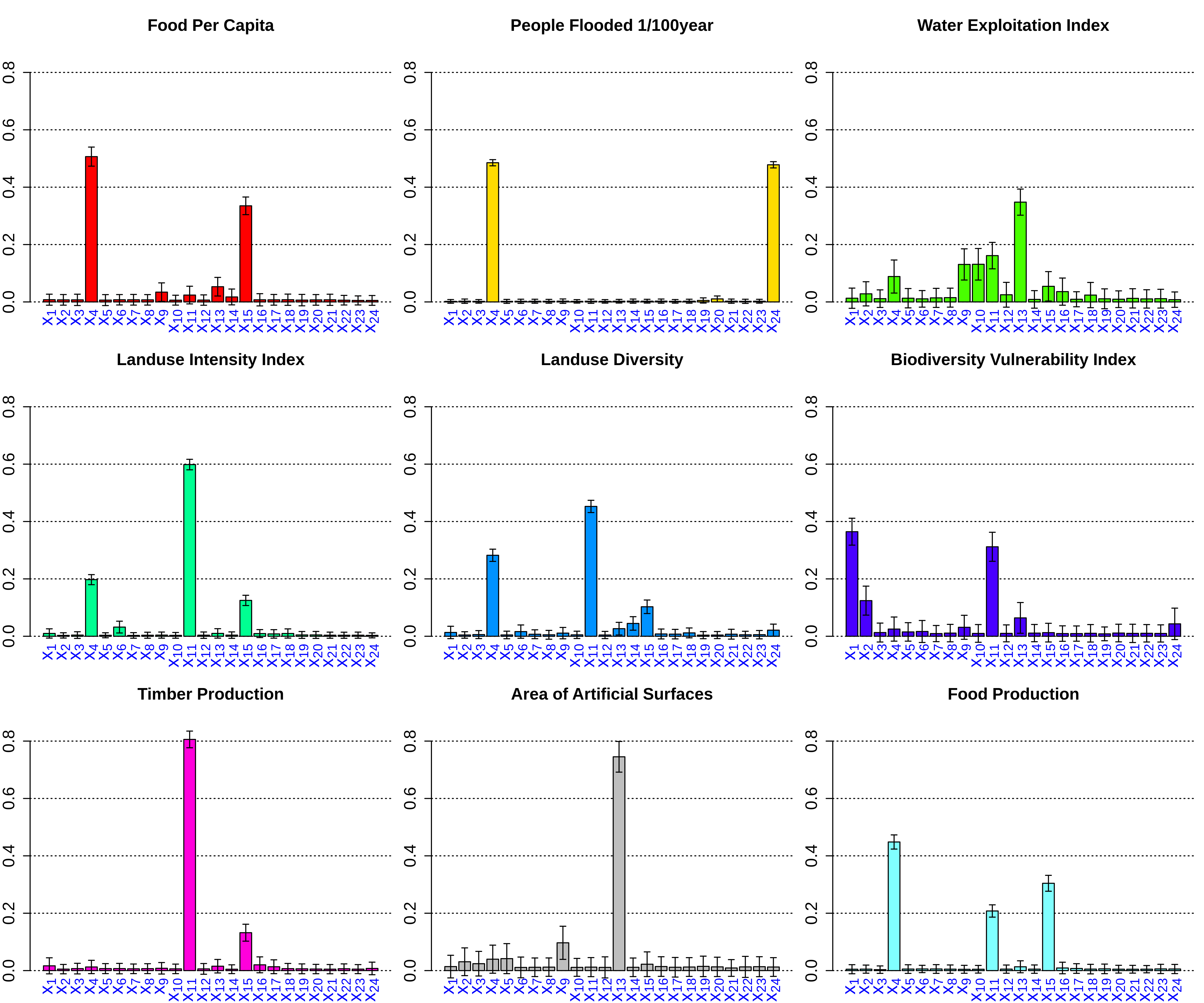}
\caption{Barplots of the posterior mean estimates of the first-order indices from the multivariate sensitivity analyses for some selected IAP2 outputs for EU-wide average data.}\label{fig5a}
\end{figure}

\begin{figure}[hbt!]
\centering
\includegraphics[width=1\linewidth]{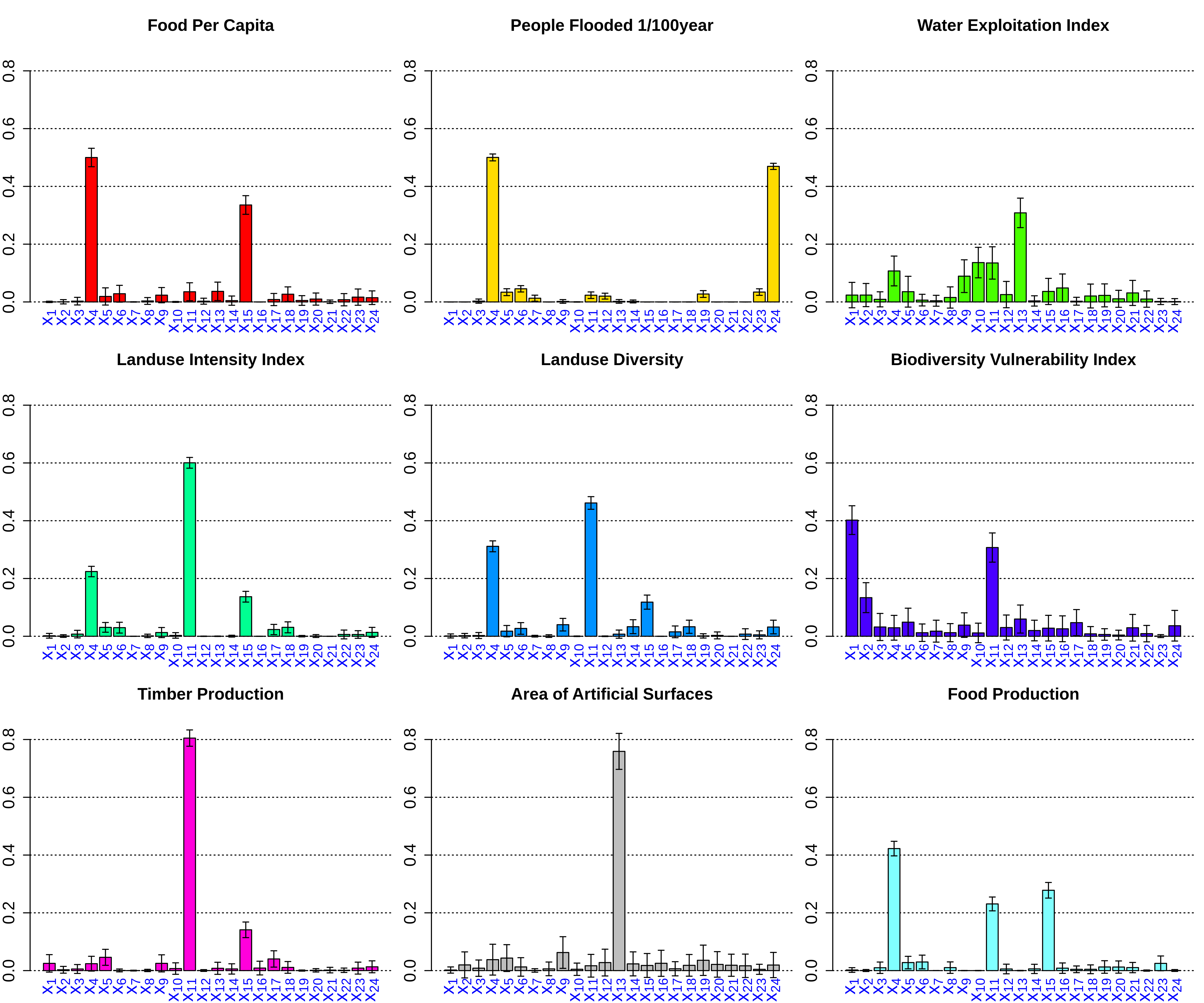}
\caption{Barplots of the posterior mean estimates of the total indices from the multivariate sensitivity analyses for some selected IAP2 outputs for EU-wide average data.}\label{fig5b}
\end{figure}%

\begin{figure}[hbt!]
\centering
\includegraphics[width=1\linewidth]{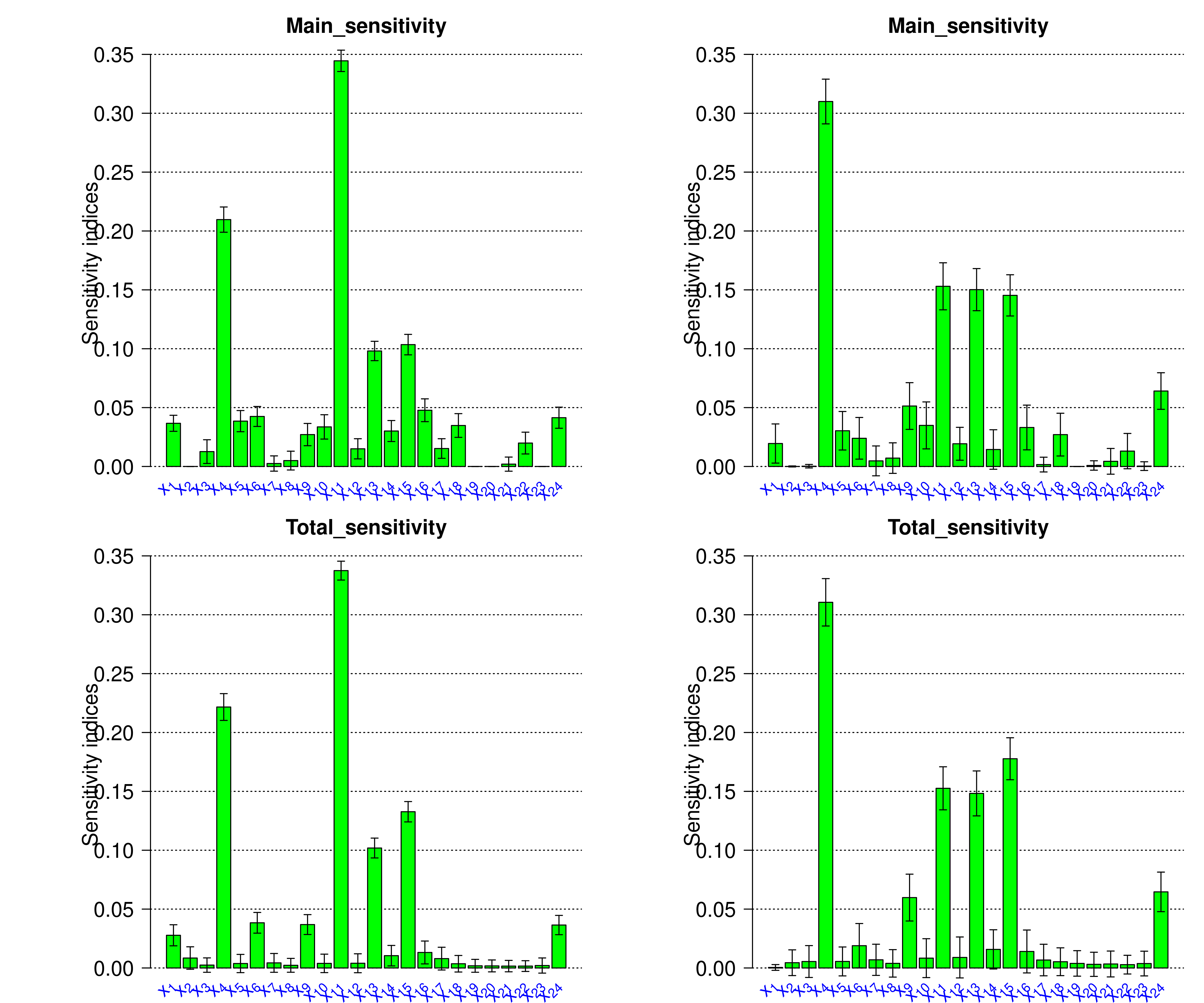}
\caption{Boxplots of the posterior mean for the multivariate generalised indices for EU-wide average data. Note: The left plots corresponds to standard approach while the right-plots are computed from vector projected method using affine transformation.}\label{fig6}
\end{figure}%

Figures~\ref{fig5a} and \ref{fig5b} are bar plots showing the posterior estimates for the first-order and total sensitivity indices for the EU-wide aggregated data for some selected outputs. The relative contribution of each input $X_j$ can be
either positive or very small negative values close to zero. This usually occurs when the posterior samples for the variances are computed using differences of expectations, especially when the Monte Carlo sample size is not large enough. Therefore, the absolute magnitudes of the sensitivity indices can be used to rank the inputs. We focus on the four largest values for the first-order indices using their posterior mean estimates. We see that the \textit{food per capita} is highly sensitive to change in food imports, population, agricultural mechanisation and agricultural yields.

The \textit{people flooded in a 1 in 100-year event} is mostly affected by the level of flood protection and population change. It is also completely nonsensitive to the other remaining variables. The \textit{water exploitation index} is affected mostly by GDP, change in agricultural yields, water-saving due to technological change and food imports. We note that change in agricultural yields, population and change in food imports are the most important variables affecting the \textit{land use intensity index, land use diversity index} and \textit{timber production}.
The emissions scenario, climate model, and socio-economic scenarios are mostly non-significant factors, except for the \textit{biodiversity vulnerability index}, which shows some degree of sensitivity to many of the inputs. This reflects its position at the bottom of the modelling chain, where it is influenced by most of the other sectoral models within the IAP2 model.

For the total sensitivity indices in Figure \ref{fig5b}, there are significant differences between the main and total indices emphasising the contribution of interactions among the inputs. We noted that the values of the main indices are clearly different from the corresponding total effect indices. Some of the sectoral outputs are still sensitive to the variables identified under the main indices as highly significant. For example, the \textit{food per capita} output is still most affected by changes in food imports and population, but it also shows sensitivity to fluvial flood events, preferences for living close to urban or rural areas, and changes in agricultural inputs to reduce diffuse pollution.

Similar to the main indices, changes in agricultural yields have the predominant influence on the \textit{land use intensity index} and \textit{land use diversity index} and \textit{timber production}, but these also show sensitivity to many other inputs, such as bioenergy crop demand and land set-aside for conservation, in the total effect indices. Other variables also show greater sensitivity to other parameters due to multiple interactions between the variables. For instance, the \textit{biodiversity vulnerability index} is now sensitive to other parameters such as changes in food imports and bioenergy production, set-aside, change in irrigation efficiency and fluvial flood events. It is interesting that it shows sensitivity to lots of parameters but that there is a large uncertainty associated with the interaction between factors. This is an indication that interaction between factors played a major role in these sets of data.

The generalised sensitivity indices based on the vector projection for multivariate outputs, which measures the comprehensive effect of the inputs on the multiple outputs, are given in Figure~\ref{fig6}. This takes into consideration the correlation between multiple outputs when computing the indices. The results are similar to those reported in Figures \ref{fig5a} and \ref{fig5b}. Clearly, the change in agricultural yields, population, GDP and food import variables are significant for both main and total indices. The less significant variables are the level of flood protection, preference for living close to urban or rural areas and the emissions scenario. We note that the change in agricultural yields dominates the given results.

\begin{figure}[!htb]
\centering
\includegraphics[width=1\linewidth]{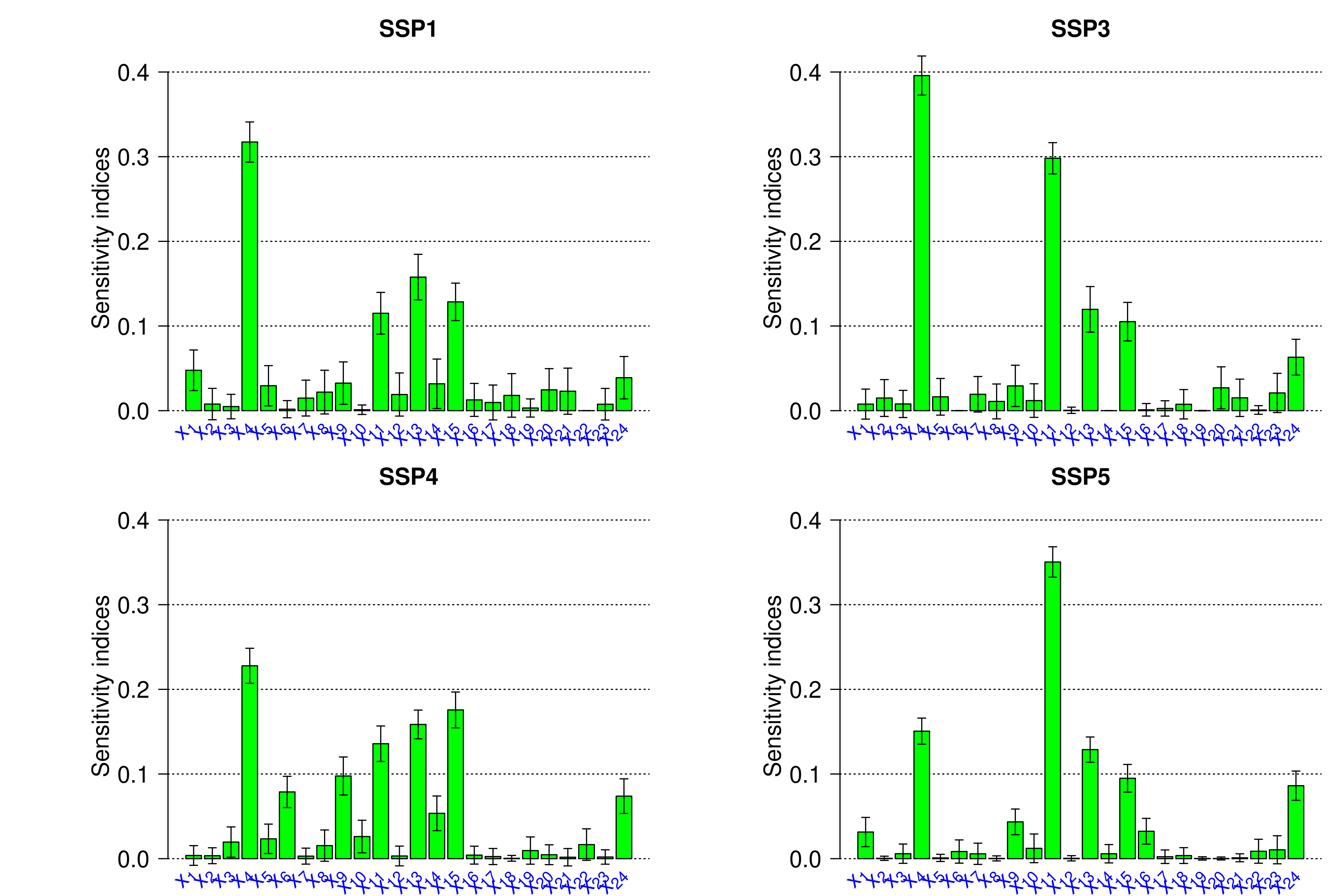}
\caption{Boxplots of the posterior mean and their uncertainty estimates for the aggregated main indices for SSP1, SSP3, SSP4 and SSP5 under the EU-wide average data.}\label{fig7}
\end{figure}

Next, we study the implications of the socio-economic scenarios on the sensitivity indices. Figure~\ref{fig7} shows the barplot for the posterior distribution of the main and total effects. There is a divergence in patterns across the scenarios. Furthermore, a different combination of variables is selected in each scenario as the essential parameters. However, the parameter change in food imports remains the most important driver across all four scenarios. The only difference is that, under SSP1, the level of flood protection is shown to be slightly more important than food imports. Apart from these two variables, other relatively significant parameters are the emissions scenario and change in agricultural mechanisation for SSP3, and the emissions scenario, change in dietary preference for beef/lamb and agricultural yields for the SSP4 scenario.

We also performed the sensitivity analysis for the regional data. The corresponding regional sensitivity indices are provided in Figure~\ref{fig8} in Appendix \ref{appA}. The results are similar across the regions to those reported in Figure~\ref{fig6}. The first four most sensitive parameters are the change in agricultural yields, GDP, food imports and population. Some variables such as water savings due to behavioural change, reducing diffuse source pollution from agriculture and protected area change are generally nonsensitive to most of the variables and thus may require less attention when parameterising the IAP2 model. This set of variables could be fixed to reduce the dimension of the problem.


\vspace{-20pt}

\section{Conclusion}
We proposed a new methodology for Bayesian global sensitivity analysis for large multivariate output data with a focus on computationally demanding models. The method was applied to a large dataset with a large number of correlated variables from the IMPRESSIONS IAP2 model. The goal was to estimate the expected outputs of such models given the distribution of inputs. These models are usually heavily parameterised, and the relative contribution of these parameters are often ignored in uncertainty quantification. The main objective of this paper was to investigate how changes in each parameter affect the sectoral and cross-sectoral indicators of the IAP2 model.

The computational cost of performing sensitivity analysis for complex models depends on the number of input variables making the approach intractable for high dimensional input data. To address the problem, we have developed an efficient strategy for computing sensitivity indices for large dimensional and multivariate data. By exploiting the sparsity of covariance matrices, using compactly supported correlation functions, we have shown how the parameters from the multivariate sparse Gaussian process model could be estimated. The procedure requires neither storing nor inverting the prohibitive full $n \times n$ covariance matrix, thus significantly reducing the computational and memory complexity. The MSGP emulator provided faster and cheaper approximation to the real data.

\citet{farah2014bayesianb, kaufman2011efficient} have earlier used this approach, but we have extended it to the case of multivariate outputs and adapted it for computing sensitivity indices for a high-dimensional problem which has not been done before. Our proposed method also took advantage of cross-correlation between the outputs to improve multivariate predictions. We also explored parallelisation in the posterior computation by distributing the likelihood estimation on different nodes using the multicore computing environment. This combined effort enabled us to achieve massive scalability in the calculation of sensitivity indices, as seen in Table \ref{sparse0}.

We applied the proposed methods on two different test cases the non-monotonic Sobol g-function \citet{saltelli2002making} and arctangent temporal function for multivariate applications \citet{auder2011classification}. Our method provides a good approximation to the sensitivity results. Next, the method was applied to the IAP2 model for the computation of main and total sensitivity indices whether or not the input is correlated. We observed that changes in agricultural yields, food imports, GDP and population have large indices and are highly significant. However, variables such as water savings due to behavioural change, reducing diffuse source pollution from agriculture, protected area change have negligible effects and could be fixed.

We have applied spatial aggregation as a strategy to reduce the data dimension. The effect of this on the MSGP and its implications for the outcomes of the sensitivity analysis is not clear. Similarly, it would be interesting to demonstrate this proposed method for computing sensitivity indices for spatiotemporal data without aggregation.
The analysis provides a thorough assessment of the effects of input uncertainty on the IAP2 model predictions. This enables a better understanding and interpretation of the complex relationships between the input drivers and the sectoral output indicators under the different socio-economic scenarios and for the five European regions. Such information is essential for further refinement and application of the IAP2 model for use in climate change adaptation planning across Europe.

\section{Acknowledgements}
The authors gratefully acknowledge the support of EPSRC and NERC grants EP/R01860X/1, EP/S00159X/1, EP/V022636/1 and NE/T004002/1.

\bibliographystyle{apalike}  
\bibliography{Manuscript}  
\clearpage

\newpage
\begin{appendices}
\section{LHS sampling details and additional Figures}\label{appA}
\begin{table}
\caption{The following are the ten emission/climate combinations used for analyses. \label{ref11}}
\centering
\small
\begin{tabular}{|l|l|}
\hline
RCPs& \\
\hline
2.6 &0 = EC-EARTH~RCA4; 1 = MPIESMLR-REMO; 2 = NorESM1-M~RCA4 \\
\hline
4.5 & 0 = HadGEM2-ES~RCA4; 1 = MPI-ESM-LR~CCLM4; 2 = GFDL-ESM2M~RCA4 \\
\hline
8.5 & 0=HadGEM2-ESRCA4; 1=CanESM2-CanRCM4;\\
&  2=IPSL-CM5A-MRWRF; 3=GFDL-ESM2\\
\hline
\end{tabular}
\end{table}
The three emission levels used in this analysis are denoted as $0 = RCP 2.6$; $1 = RCP 4.5$; $2 = RCP 8.5$. For all the three emissions we have a total of ten climate models (ie, RCP 2.6, 4.5 and 8.5 have 3, 3 and 4 climate models, respectively.) The following ten climate/emission combinations used for the analyses are given in the Table~\ref{ref11} below.
We also have five socio-economic scenarios (SSPs) denoted as 0 = "SSP1 / We are the world"; 1 = "SSP3 / Icarus"; 2 =  "SSP5 / Should I Stay Or Should I Go"; 3 = "SSP4 / Riders on the Storm"; 4 = "Baseline" and three time slices. Together, we have performed 30,000 simulation runs $(2000\times3\times5)$. To train the MSGP emulators, we subsample 600 data points randomly from 2000 emission/climate combinations for each time slice and SSPs, making 9000 samples $(600\times3\times5)$. Each sample represents a matrix of $\by_{n0,m}$ spatial outputs with $n0=23871$ grid points and $m=18$ which is the dimension of output data. To further pre-process the data, we have averaged the data EU-wide and regionally. The output data for emulation is denoted as $\bY_{9000 \times 18}$ for each of the six averages (EU-wide, five regions). This produces an array of $9000 \times 18 \times 6$ output data.

\begin{table}
\caption{IMPRESSIONS IAP2 parameter ranges for the four socio-economic scenarios from 2020s to 2080s. The 20 continuous variables were sample from uniform distribution. \label{ref1}}
\centering
\begin{tabular}{|l|l|llll|}
\hline
&Variables&SSP1& SSP3&SSP5&SSP4\\
\hline
X$^{\star}_1$&Emission& \{0, 1, 2\} &\{0, 1, 2\}&\{0, 1, 2\}&\{0, 1, 2\}\\
\hline
X$^{\star}_2$&Model&\{0,1, 2, 3\} &\{0, 1, 2, 3\}&\{0, 1, 2, 3\}&\{0, 1, 2, 3\}\\
\hline
X$^{\star}_3$&SSPs&\{0\} &\{1\}&\{2\}&\{3\}\\
\hline
X$_4$&Population&[3, 16] &[-37, 1]&[5, 69]&[-18, 3]\\
\hline
X$_5$&GreenRed&[3, 5] &[2, 5]&[3, 5]&[1, 5]\\
\hline
X$_6$&RumLivesDem&[-81, -5] &[-27, 38]&[-10, 94]&[-25, 39]\\
\hline
X$_7$&NonRumiLivesDem&[-33, 5] &[2, 54]&[29, 97]&[4, 54]\\
\hline
X$_8$&StructChange&[-5, 55] &[-12, 13]&[-24, 14]&[-15, 24]\\
\hline
X$_9$&TechFactor&[5, 125] &[-30, 1]&[5, 125]&[4, 125]\\
\hline
X$_{10}$&TechChange&[-10, 77] &[-10, 65]&[2, 87]&[-10, 65]\\
\hline
X$_{11}$&YieldFactor&[-22, 0.7] &[-34, -4]&[7, 105]&[5, 106]\\
\hline
X$_{12}$&IrrigEffFactor&[4, 66] &[-37, -3]&[4, 66]&[4, 66]\\
\hline
X$_{13}$&GDP&[9, 427] &[17, 66]&[21, 1312]&[19, 309]\\
\hline
X$_{14}$&CostsFactor& [90, 253] &[108, 467]&[59, 110]&[143, 452]\\
\hline
X$_{15}$&ImportFactor&[-8, 3] &[-6, 1]&[.4, 15]&[1.1, 8.2]\\
\hline
X$_{16}$&BioEneCropDemand&[.7, 18] &[1.8, 24]&[1, 21]&[1, 18]\\
\hline
X$_{17}$&ArableConservLand&[3, 6.3] &[2.2, 3]&[0.0 ,2.2]&[3, 6.3]\\
\hline
X$_{18}$&CropInputsFactor&[1, 2] &[.93, 1]&[.93, 1.2]&[.93, 1.2]\\
\hline
X$_{19}$&Coastal&[11, 500] &[13, 500]&[12, 500]&[14, 500]\\
\hline
X$_{20}$&Fluvial&[11, 500] &[13, 500]&[11, 500]&[9, 500]\\
\hline
X$_{21}$&HabitatRecreation&[-100,100] &[-100,100]&[-100,100]&[-100,100]\\
\hline
X$_{22}$&PA Forestry&[0,100] &[0,100]&[0,100]&[0,100]\\
\hline
X$_{23}$&PA Agriculture&[0,100] &[0,100]&[0,100]&[0,100]\\
X$^{\star}_{24}$&FloodProtection&\{0,1\} &\{0,1\}&\{0,1\}&\{0,1\}\\
\hline
\end{tabular}
\end{table}

\begin{figure}[!htb]
\begin{center}
\includegraphics[width=.95\textwidth]{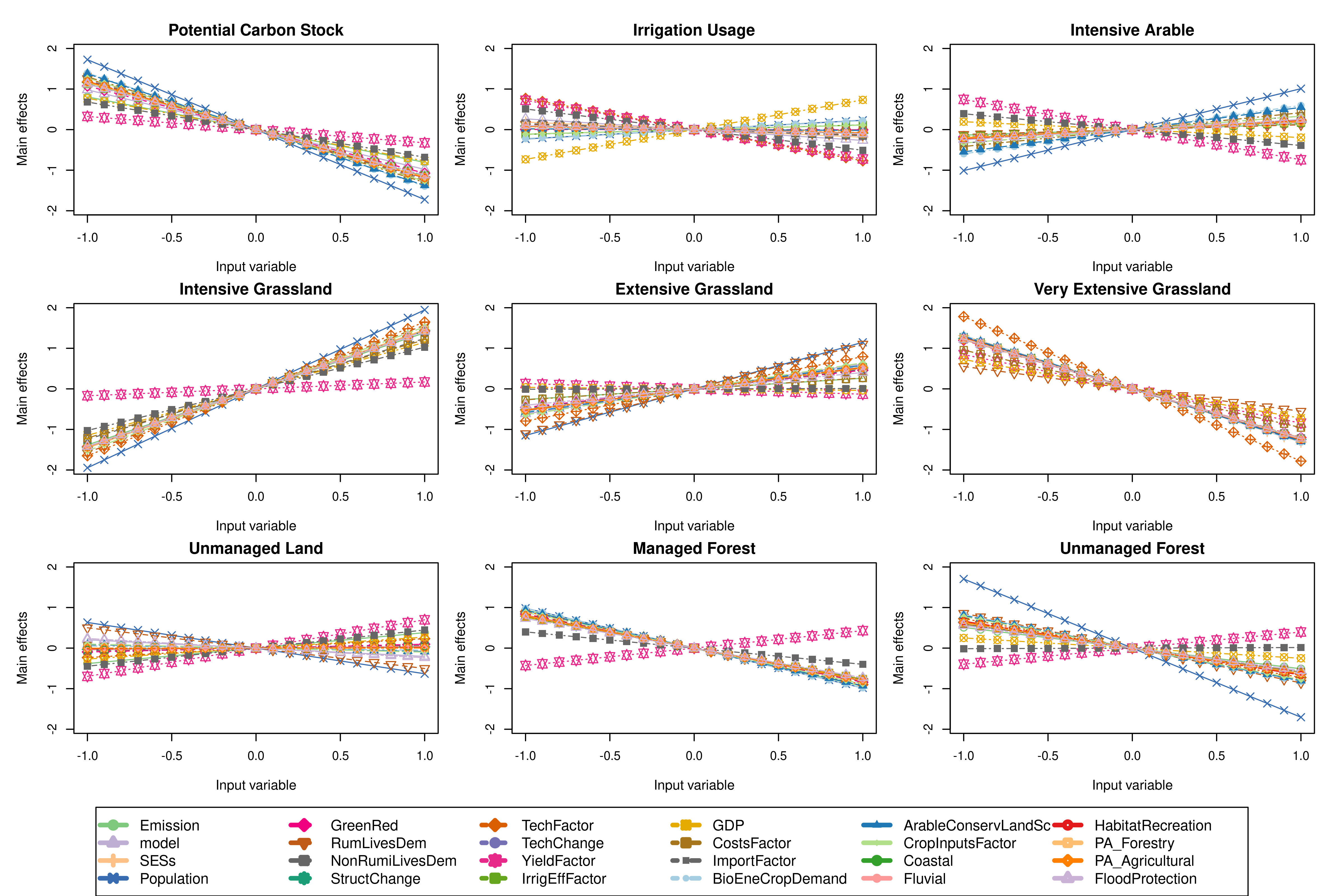}
\end{center}
\caption{Main effect plots for EU-wide aggregated data; simulation (black) and emulator predictions (green) with 95\% C.I. (red). Note: the outputs have been normalised to have a zero mean and unit variance.}\label{fig4aa2}
\end{figure}

\begin{figure}[!htb]
\begin{minipage}{.9\textwidth}
\centering
\includegraphics[width=.7\linewidth]{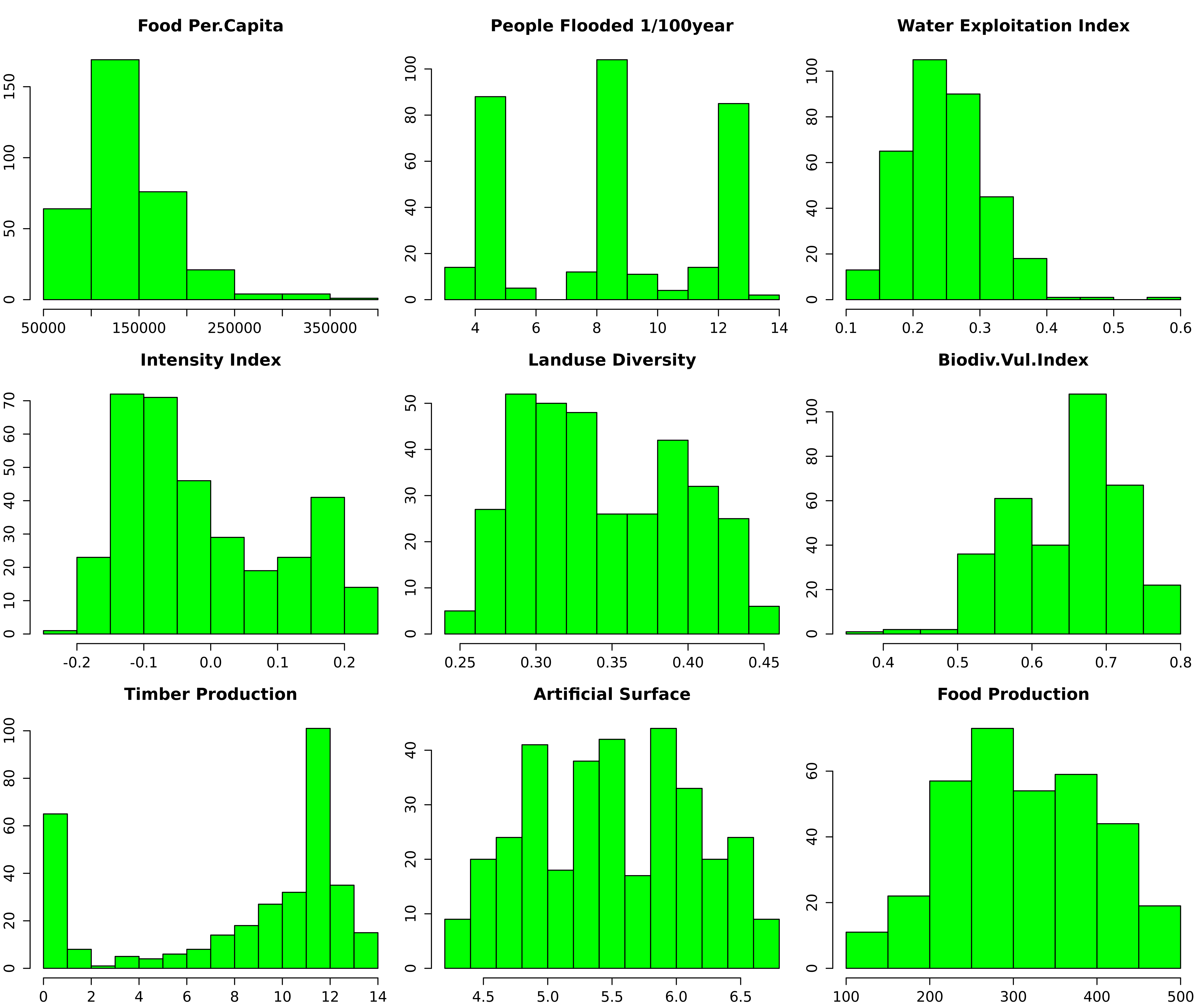}
\end{minipage}%
\\
\begin{minipage}{.9\textwidth}
\centering
\includegraphics[width=.7\linewidth]{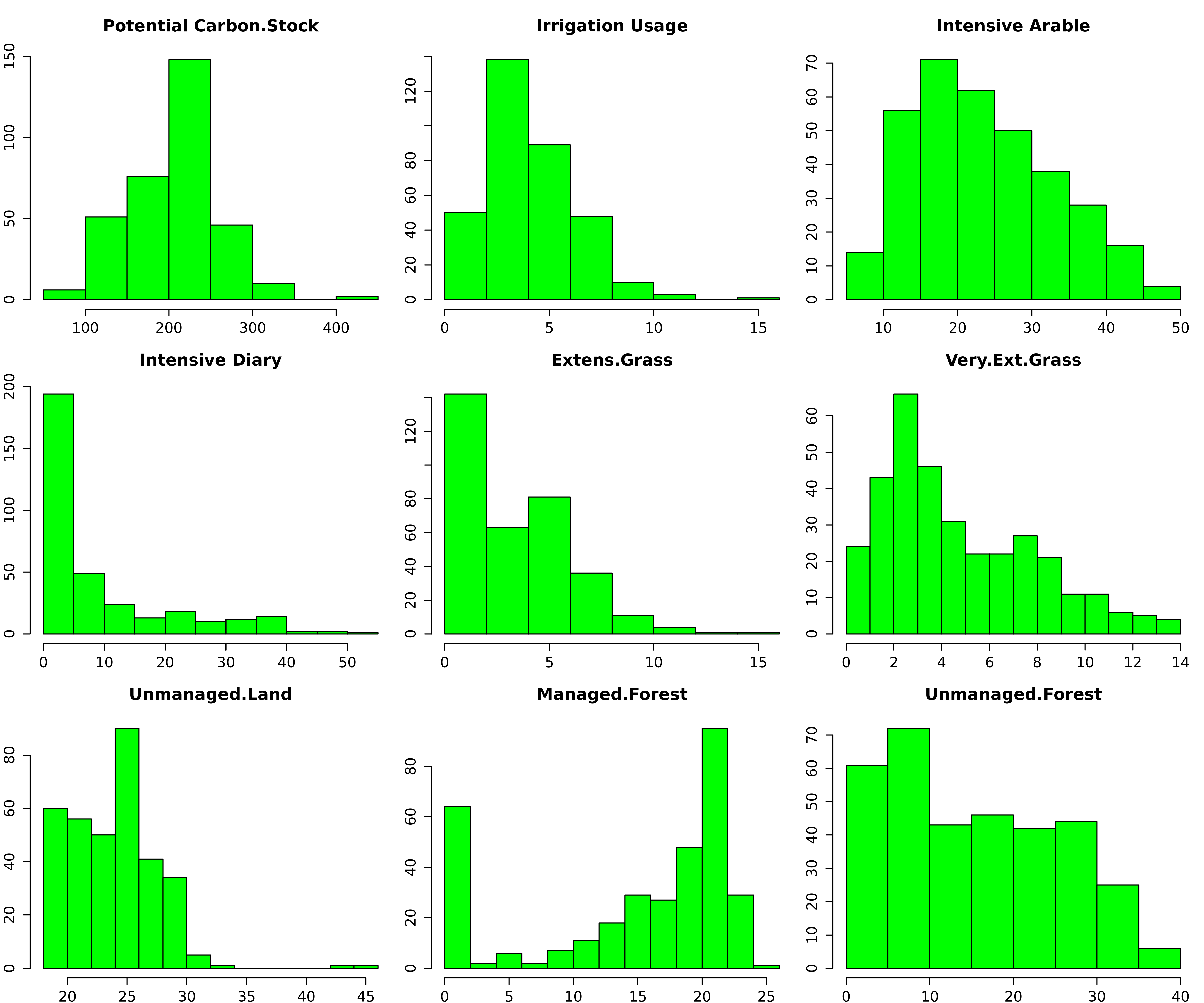}
\end{minipage}
\caption{Histograms of spatially aggregated IAP2 simulation for the 18 selected outputs, averaged across the three emission scenarios and ten climate models for SSP4.}\label{fig2b}
\end{figure}

\begin{figure}[!htb]
\centering
\includegraphics[width=.9\linewidth]{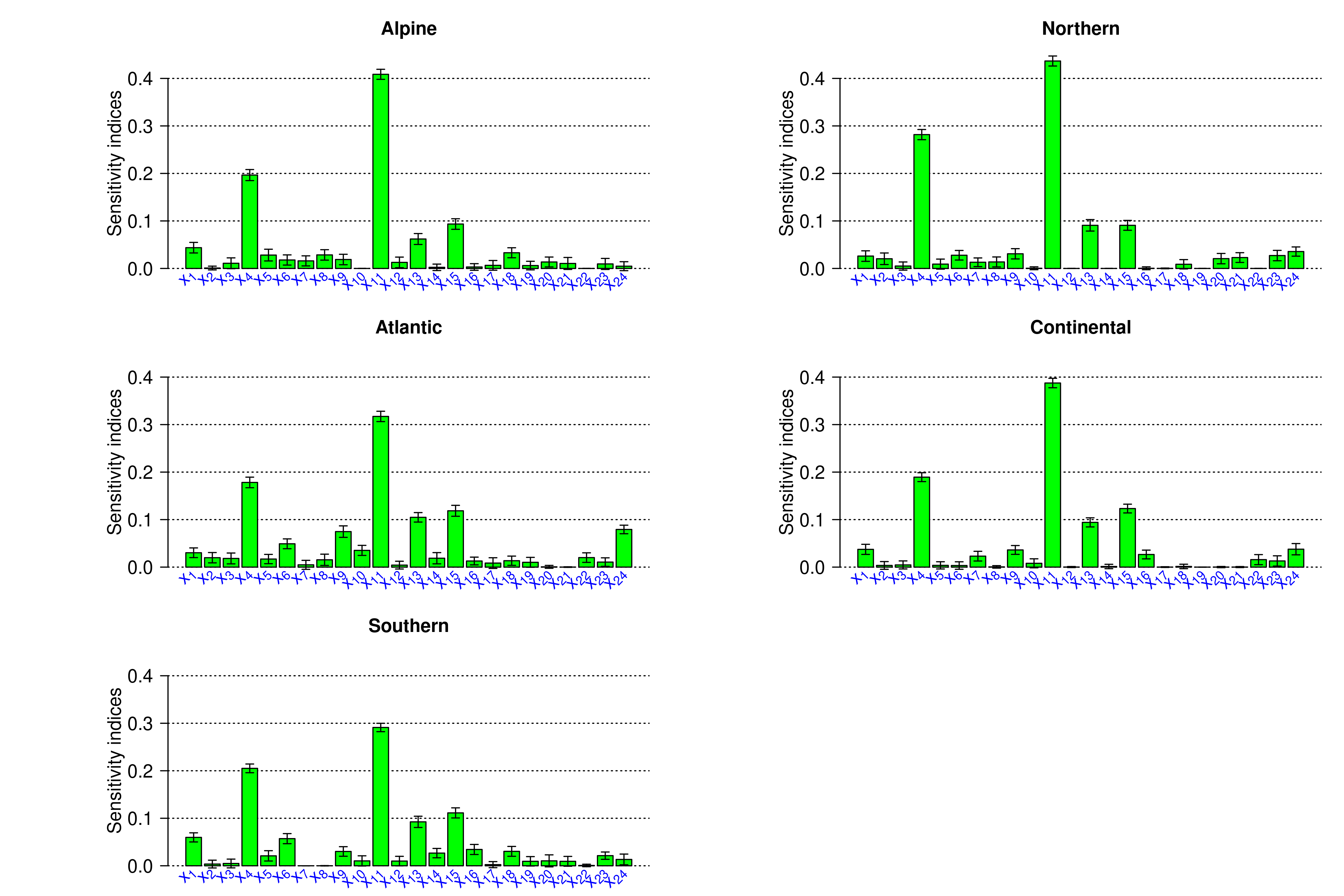}
\caption{Boxplots of the posterior mean and their uncertainty estimates for the aggregated main indices for Alpine, Northern, Atlantic, Continental and Southern regional averaged data.}\label{fig8}
\end{figure}

\section{MCMC sampling details}\label{appB}
We follow the procedure below to generate samples from the posterior distribution $p(\btau |\bY)$ and make a prediction at a new location. %
To sample $\btau$, we implement the adaptive Metropolis algorithm of \citet{vihola2012robust} to speed up mixing of the target distribution.
Suppose the value of tuning parameter $\gamma \in (0,1]$ and given a fixed target acceptance rate $\alpha^{\star}$ and symmetric proposal density $w$ such that
$w(\bx)=\hat{w}(\|\bx\|)$ for all $\bx \in \mathbb{R}^{d}$ and $\Upsilon_{1} \in \mathbb{R}^{d \times d}$ be a lower-diagonal matrix with positive diagonal elements and $\left\{\omega_{n}\right\}_{n \geq 1} \subset(0,1]$ . Then at $j^{th}$ iteration of the MCMC:

\begin{itemize}
\item Compute $\btau^{(`)}=\btau^{(j-1)}+\Upsilon_{j-1} \bu_{j}$, where $\bu_{j} \sim w$ is an independent random vector

\item Accept the proposal with probability $\alpha_{j} :=\min \left\{1, \frac{p\left(\btau^{\prime}\right) p(\bY | \btau)}{\left. p\left(\btau_{j-1}\right) p\left(\bY | \btau_{j-1}\right)\right\}}\right.$

\item If the proposal $\btau^{\prime}$ is accepted, set $\btau_{j}=\btau^{\prime}$. Otherwise, $\operatorname{set} \btau_{j}=\btau_{j-1}$

\item Compute the Cholesky factor $\Upsilon_j$ that satisfy\\
$\Upsilon_{j} \Upsilon_{j}^{T}=\Upsilon_{j-1}\left(I+\omega_{j}\left(\alpha_{j}-\alpha_{*}\right) \frac{\bu_{j} \bu_{j}^{T}}{\left\|\bu_{j}\right\|^{2}}\right) \Upsilon_{j-1}^{T}$, where $\Upsilon_{j}$ is the lower-diagonal matrix with positive diagonal elements and $I \in \mathbb{R}^{d \times d}$ stands for an identity matrix.
\end{itemize}

Once the posterior samples from $\btau^{(j)}$ has been drawn. We can easily obtain the predictive distribution of $Y^{j}$ given $\bh(\bx_0)$ and $\bH(\bx)$. Recall that that the joint conditional distribution of $\bY(\bx_{0})$ at a new location $\bx_0$ and $\bY(\bx) $ is given as a matrix t- distribution $p(\bY_0|\bY, \btau) \sim MT(\bQ, \bS,\mathfrak{D},\delta)$ with location matrix $\bQ$ and scale matrix $\mathfrak{D}$.
Suppose further that $\mathbf{q}_{i}$ is the $i^{th}$ row of $\bQ$ and $q_{i s}$ is the $i^{th}$ element of $\bQ$, $\mathfrak{S}_{i i}$ is the $i^{th}$ diagonal element of $\mathfrak{R}$ and $\bS_{s s}$ is the $s^{th}$ diagonal element of $\bS$, then the marginal posterior predictive distributions at the $i^{th}$ simulator point is given as $\mathbf{y}_{0 i}=f\left(\mathbf{x}_{0 i}\right),$ and the $s^{th}$ output, $y_{0, i s}=f_{s}\left(\mathbf{x}_{0 i}\right)$ are multivariate and univariate $t$ distributions, respectively:

\begin{equation}
\mathbf{y}_{0 i}| \bY, \mathbf{\btau} \sim \mathrm{t}_{q}\left(\mathbf{q}_{i}^{T}, \frac{\mathfrak{S}_{ii} \bS}{\hat{\delta}}, \hat{\delta}\right) , \quad
y_{0, i s}| Y, \mathbf{\btau} \sim \mathrm{t}\left(q_{i s}, \frac{\mathfrak{S}_{ii} \bS_{s s}}{\hat{\delta}}, \hat{\delta}\right)
\end{equation}

\subsection{Pseudo-code for MCMC Algorithm}\label{algo1}
\begin{itemize}
\item [1] Substitute $R(x,x')=\prod\limits_{k=1}^{p} \exp \left\{-\bphi_{k}\left|\bx_{.,k}-\bx_{.,k}^{\prime}\right|^{2}\right\} \quad$ (Power exponential) with either truncated power or Bohman correlation function $R(t ; \btau)$) given above
\begin{itemize}
\item (a) Identify two points $x_i$ and $x_j$ in sample space such that $t=|\bx_{ik} - \bx_{jk}| < \btau_k$, $\quad \forall k$.
\item (b) Create a sparse representation of the $n \times n$ correlation matrix $\Re(\btau)$ with only these pairs while the remaining pair will contribute zero to the correlation matrix to complete the model specification
\item (c) Store only the non-zero off-diagonal elements with their row and column indices
\item (d) Parallelise steps (a-c) above using the Rcpp and Armadillo linear Algebra packages
\end{itemize}
\item [2] Assign a matrix-normally inverse-Wishart (MNIW) prior distribution to $\bB$ as $p\left(\boldsymbol{\bB} | \boldsymbol{\Sigma},\phi\right) \sim N\left(\boldsymbol{\bB}_{0}, \boldsymbol{\Sigma} \otimes \bLambda_{0}\right)$
and $\bSigma$ as $p\left(\boldsymbol{\Sigma}|\bphi\right) \sim \mathcal{W}^{-1}\left(\mathbf{\bS}_{0}, \delta_{0}\right)$ where $\bB_0$, $\Lambda_0$, $\bS_0$ and $\delta_0$ are given prior hyperparameters
\item [3] Obtain the marginal posterior distributions for $\bB$ and $\bSigma$ using Gibbs sampling
\item [4] Generate a sample of the posterior $p(\btau|\bD)$ using a robust adaptive Metropolis-within-Gibbs step, for each $\bB$ and $\Sigma$
\item [5] Solve the sparse linear systems, such as $\Re^{-1}\left(\mathbf{D}-\mathbf{H} \boldsymbol{\bB}\right)$, that occur in the posterior density
\item [6] Repeat steps (1-5) $N$ times to obtain $N$ iterations of MCMC samples for $\widehat \bTheta= \{\bB, \bSigma, \btau\}$.
\end{itemize}

\section{Posterior expectations for computing the sensitivity indices}\label{appC}
The formula required for computing all the expectations for $S_{j}$ and $S_{j}^{T}$, that is, $\operatorname{cov}(Y)=E\left(Y^{2}\right)-$ $(E(Y))^{2}, E^*\left(\left(E\left(Y \mid x_{j}\right)\right)^{2}\right),$ and $E\left(\left(E\left(Y \mid x_{\sim j}\right)\right)^{2}\right) $ are given below:
The expectation and covariance of $Y$ are given by
$$ E(Y)=\int_{\bx} f(\bx) \prod_{j=1}^{p} d G_{j}\left(x_{j}\right)$$
$$\operatorname{var}(Y)=\int_{\bx} f^{2}(\boldsymbol{\bx}) \prod_{j=1}^{p}dG_{j}\left(x_{j}\right)-(E(Y))^{2}.$$ Define
\begin{equation}\label{exp1}
E\left(Y \mid x_{j}\right)=\int_{\left\{x_{j'}: j' \neq j\right\}} f\left(x_{1}, \ldots, x_{j}, \ldots, x_{p}\right) \prod_{\{j': j' \neq j\}} dG_{j'}\left(x_{j'}\right).
\end{equation}

For each specified value $x_{j}$ of the $j^{th}$ input, squaring the expression for $E(Y |x_j )$ in \eqref{exp1} and taking its expectation, we have

\begin{multline}
E^*\left(\left(E\left(Y \mid \bx_{j}\right)\right)^{2}\right)=\\
\int\left\{\int_{\left\{x_{j'}: j' \neq j\right\}} \bff\left(x_{1}, \ldots, x_{j}, \ldots, x_{p}\right) \prod_{\{j': j' \neq j\}} dG_{j'}\left(x_{j'}\right)\right\}^{2}
\quad \times dG_{j}\left(x_{j}\right) \\
=\int\left\{\int_{\left\{x_{j'}: j' \neq j\right\}} \int_{\left\{x_{j'}^{\prime}: j' \neq j\right\}}\right. \\
\quad f\left(x_{1}, \ldots, x_{j}, \ldots, x_{p}\right) f\left(x_{1}^{\prime}, \ldots, x_{j}, \ldots, x_{p}^{\prime}\right)
\left.\prod_{\{j': j' \neq j\}} 
  dG_{j'}\left(x_{j'}\right) \prod_{\{j': j' \neq j\}} dG_{j'}\left(x_{j'}^{\prime}\right)\right\}\\
  \times dG_{j}\left(x_{j}\right) \\
=\int_{\bx} \int_{\left\{x_{j'}^{\prime}: j' \neq j\right\}} f(\bx) f\left(x_{1}^{\prime}, \ldots, x_{j}, \ldots, x_{p}^{\prime}\right)
\prod_{j'=1}^{p} dG_{j'}\left(x_{j'}\right) \prod_{\{j': j' \neq j\}} dG_{j'}\left(x_{j'}^{\prime}\right).
\end{multline}
Similarly, for the expectations required for the total sensitivity indices,
let $$\boldsymbol{x}_{\sim j}=\left(x_{1}, \ldots, x_{j-1}, x_{j+1}, \ldots, x_{p}\right),$$
then
\begin{equation}
E\left(Y \mid \boldsymbol{x}_{\sim j}\right)=\int f\left(x_{j}, \boldsymbol{x}_{\sim j}\right) dG_{j}\left(x_{j}\right),
\end{equation}
and, analogously to the derivation above. Similarly, we can write
\begin{equation}
E^*\left(\left(E\left(Y \mid \boldsymbol{x}_{\sim j}\right)\right)^{2}\right) \\
=\iint f\left(x_{j}, \boldsymbol{x}_{\sim j}\right) f\left(x_{j}^{\prime}, \boldsymbol{x}_{\sim j}\right) dG_{j}\left(x_{j}^{\prime}\right) \prod_{j'=1}^{p} dG_{j'}\left(x_{j'}\right).
\end{equation}

\subsection{Algorithm summary 2}
We have summarised the algorithm to compute the main and total sensitivity indices for the multivariate outputs:
\begin{itemize}
\item [1] Normalize $\bY=\left(Y^{(1)}, \ldots, Y^{(m)}\right)$ to common scale by dividing the centered column of $\bY$ by their standard deviations to obtain $\bY'=\frac{\left(\bY-\bar \bY\right)} {\sigma(\bY)}$, where $\sigma$ is the standard deviation.
\item [2] Develop a fast approximation of a continuous function $\hat \bY = f(\bx)$ using the multivariate sparse Gaussian process (MSGP) model based on the marginal samples of the inputs and corresponding set of simulated multivariate outputs.
\item [3] Compute the following functions $\left\{f_{j}^{(k)}, f_{\sim j}^{(k)}, f_{j, \sim j}^{(k)}\right\}_{k=1}^{m}$ of multivariate outputs using the posterior predictive predictions given in the equation (\ref{md2}).

\item [4] Estimate main and total sensitivity indices of the input variables for noncorrelated multivariate outputs given in equation~(\ref{mu3}) from their full marginal contributions $$\left\{\bOmega_{j}^{(k)}, \bOmega_{\sim j}^{(k)}, \bOmega_{j, \sim j}^{(k)}\right\}_{k=1}^{m}$$.

\item [5] Obtain the empirical estimate of the correlation coefficient matrix $R_{m,m}(\bY)$ from the training data $\bD$. Calculate the projected main and total sensitivity indices of the input variables for correlated multivariate outputs given in equations~(\ref{mu4} and \ref{mu5}) from their full marginal contributions $$\left\{\bOmega_{j}^{(k)}, \bOmega_{\sim j}^{(k)}, \bOmega_{j, \sim j}^{(k)}\right\}_{k=1}^{m}.$$ 
\end{itemize}

\begin{equation}\label{app3}
\bOmega_{j}=\left[\begin{array}{cccc}
\operatorname{Cov}\left(f_{j}^{(1)}, \bff^{(1)}\right) & \operatorname{Cov}\left(f_{j}^{(1)}, \bff^{(2)}\right) & \cdots & \operatorname{Cov}\left(f_{j}^{(1)}, \bff^{(m)}\right) \\
\operatorname{Cov}\left(f_{j}^{(2)}, \bff^{(1)}\right) & \operatorname{Cov}\left(f_{j}^{(2)}, \bff^{(2)}\right) & \cdots & \operatorname{Cov}\left(f_{j}^{(2)}, \bff^{(m)}\right) \\
\vdots & \vdots & \ddots & \vdots \\
\operatorname{Cov}\left(f_{j}^{(m)}, \bff^{(1)}\right) & \operatorname{Cov}\left(f_{j}^{(m)}, \bff^{(2)}\right) & \cdots & \operatorname{Cov}\left(f_{j}^{(m)}, \bff^{(m)}\right)
\end{array}\right]\end{equation}

Note: $\bOmega_{\sim j}$ are given analogously with $''j''$ replaced with $''\sim j''$.

\begin{equation}\label{app4}
\bOmega_{j, \sim j}=\left[\begin{array}{ccc}
\operatorname{Cov}\left(f_{j, \sim j}^{(1)}, \bff^{(1)}\right) & \cdots & \operatorname{Cov}\left(f_{j, \sim j}^{(1)}, \bff^{(m)}\right) \\
\vdots & \ddots & \vdots \\
\operatorname{Cov}\left(f_{j, \sim j}^{(m)}, \bff^{(1)}\right) & \cdots & \operatorname{Cov}\left(f_{j, \sim j}^{(m)}, \bff^{(m)}\right).
\end{array}\right]
\end{equation}

The interaction sensitivity index $\bP_{j, \sim j}$ can be defined by normalizing $\bQ_{j, \sim j}$ with $\|\bOmega\|$ as
$$\bP_{j, \sim j}=\frac{\bQ_{j, \sim j}}{\|\bOmega\|}=\frac{\left\langle \bOmega_{j, \sim j}, \bOmega(\bY)\right\rangle}{\|\bOmega\|^{2}}.$$
The total sensitivity index $\bP^T$ based on the vector projection can be defined by
\begin{equation}\label{mu5}
\bP^T = \bP_{j, \sim j} + \bP_{j} = \frac{\left\langle (\bOmega_{j, \sim j}+\bOmega_{j}), \bOmega(\bY)\right\rangle}{\|\bOmega(\bY)\|^{2}}.
\end{equation}

\section{Covariance tapering and Schur product theorem}\label{appD}
We also consider the product of Matern and Wendland correlation functions given below.
Wendland's compactly supported functions belong to the family of compactly supported radial functions. It can be constructed by a repeated operation on the truncated power function (which we know to be strictly positive definite and radial on $\mathbb{R}^{s}$ for $\left.s \leq 2 \ell-1\right)$.

\begin{equation}
\varphi_{s, 2}(t) \doteq(1-t)_{+}^{\ell+2}\left[\left(\ell^{2}+4 \ell+3\right) t^{2}+(3 \ell+6) t+3\right]
\end{equation}
where $\ell=\lfloor s / 2\rfloor+k+1$, $t=\left\|\bx -\bx' \right\|$  and the symbol $\doteq$ denotes equality up to a multiplicative positive constant.
\begin{equation}\label{matern}
R\left(\mathbf{x}_{i}, \mathbf{x}_{j}\right)= C_{\phi,\nu}(|\mathbf{x}_{i}-\mathbf{x}_{j}|) = \frac{1}{2^{\nu-1} \Gamma(\nu)}\left(\left\|\mathbf{x}_{i}-\mathbf{x}_{j}\right\| \phi\right)^{\nu} \mathcal{K}_{\nu}\left(\left\|\mathbf{x}_{i}-\mathbf{x}_{j}\right\| \phi\right) ; \phi>0, \nu>0. 
\end{equation}
Here $\Gamma$ is the Gamma function, and $\mathcal{K}_{\nu}$ is the modified Bessel function of the second kind of order $\nu$. The parameters $\phi$ is related to the effective range. For certain values of $\nu$, the Matern covariance function has appealing forms. For example, if $\nu=0.5, C_{\phi, \nu}$ becomes an exponential covariance, if $\nu=n_0+0.5$ with $n_0$ an integer, $C_{\phi, \nu}$ becomes the product of an exponential covariance and a polynomial of order $n$.

Tapering involves multiplying the data covariance matrix element-wise by a sparse correlation matrix.
In both cases, one must maintain the positive definiteness of any sparse matrices.
Let $C_{\btau}$ be a correlation function that is identically zero outside a particular range described by $\btau$. Now consider a tapered correlation that is the Schur or Hadamard product of $C_{\btau}$ and $C$
\begin{equation}\label{tap1}
C_{\text {tap }}\left(\mathbf{\bx}, \mathbf{\bx}^{'}; \tau \right)=C\left(\mathbf{\bx}, \mathbf{\bx}^{'}\right) C_{\btau}\left(\mathbf{\bx}, \mathbf{\bx}^{'}\right).
\end{equation}

An approximation to the original process is obtained by replacing the correlation matrices in \ref{tap1} based on $C$ by those defined by $C_{\text {tap }}$ to model dependence. The assumption is that the product $C_{\text {tap }}$ of correlation matrices preserves some of the shapes of original correlation $C$ which is identically zero outside the chosen range $\tau$ thus making $C_{\text {tap }}$ a valid correlation, by using the Schur product theorem the tapered
covariance function is positive definite and thus a valid covariance function \citep{furrer2006covariance,horn2012matrix}. See the proof of the Schur product theorem below.

\subsection{Schur theorem}
If $A$ and $B$ are positive semidefinite $n \times n$ matrices, then so is $A \circ B$. If, in addition, $B$ is positive definite, and $A$ has no diagonal entry equal to $0$, then $A \circ B$ is positive definite. In particular, if both $A$ and $B$ are positive definite, then so is $A \circ B$.

Definition: Two $m \times n$ matrices, $A$ and $B$, have Schur product $A \circ B = \{a_{ij} b_{ij}\}$.\\
Proof:\\
Consider an integral operator $C(f) =\int_a^b C(\bx, \bx') f(\bx) d\bx$
such that $f \in [a,b]^R$ and covariance function $H(\bx, \bx')$ having the same condition, and consider the pointwise product kernel such that $L(\bx, \bx') = C(\bx, \bx') H(\bx, \bx')$, and the associated integral operator then,
\begin{equation}
L(f) = \int_a^b L(\bx, \bx') f(\bx) d\bx = \int_a^b C(\bx, \bx') H(\bx, \bx') f(\bx) d\bx.
\end{equation}
The linear mapping $f \rightarrow C(f)$ is a limit of matrix-vector multiplications (approximate the integral as a finite Riemann sum), and many properties of integral operators can be obtained by taking appropriate limits of results known for matrices. Therefore, the pointwise product of integral kernels produces an integral operator that is a continuous analogue of the Hadamard product of matrices.

Suppose, an integral kernel $C(\bx, \bx')$ has the property that $$\int_a^b \int_a^b C(\bx, \bx') f(\bx) \bar f(\bx')d\bx d\bx' \ge 0, $$
for all $f \in [a, b]^R$, then $C(\bx, \bx')$ is said to be a positive semidefinite kernel. This is a standard result from (Mercer's theorem) that if $C(\bx, \bx')$ is a continuous positive semidefinite kernel on a finite interval $[a,b]$, then there exist positive real numbers $\lambda_1, \lambda_2.....$ (eigenvalues) and continuous functions $\zeta_1(\bx), \zeta_2(\bx)......$ (eigenfunctions)
such that $$C(\bx, \bx') = \sum_{i=1}^{\infty} \frac{\zeta_i(\bx), \bar \zeta_i(\bx')}{\lambda_i},$$
on $[a,b] \times [a, b]$ and the series converges absolutely and uniformly.
Therefore, if $C(\bx, \bx')$ and $H(\bx, \bx')$ are both continuous positive semidefinite kernel s on $[a, b]$, then $L(\bx, \bx')$ also has an absolute and uniform convergence such that

$$L(\bx, \bx') = \sum_{i=1}^{\infty} \frac{\psi_i(\bx), \bar \psi_i(\bx')}{\mu_i},$$
on $[a,b] \times [a,b]$ with all $\mu_i > 0$. Therefore, pointwise product kernel $L(\bx,\bx') = C(\bx, \bx') H(\bx, \bx')$ also has the representation
$$L(\bx, \bx') = \sum_{i,j=1}^{\infty} \frac{\zeta_i(\bx) \psi_j(\bx), \br \zeta_i(\bx') \bar \psi_j(\bx')}{\lambda_i \mu_j},$$
on $[a, b] \times [a, b]$ which also converges absolutely and uniformly. Then
\begin{equation}
\int_a^b \int_a^b C(\bx, \bx') f(\bx) \bar f(\bx') d\bx d\bx' = \sum_{i,j=1}^{\infty} \frac{1}{\lambda_i \mu_j} |\int_a^b \zeta_i(\bx) \psi_i(\bx) f(\bx) d\bx|^2 \ge 0.
\end{equation}
Therefore, $L(\bx,\bx')$ is also positive semidefinite.

Theorem: Let $A$ and $B \in R^n$ be positive semidefinite.
\begin{itemize}
\item $A \circ B$ is positive semidefinite.
\item if $A$ is positive definite and every main diagonal entry of $B$ is positive, then $A \circ B$ is postive definite.
\item if both $A$ and $B$ are positive definite, then $A \circ B$ is positive definite.
\item Commutivity. Unlike the standard matrix product, $A \circ B= B \circ A$.
\item Trace. For square matrices $A$, $B$, and $C$, with $B$ symmetric, $tr(A \circ B)C = tr A(B \circ C)$.
\end{itemize}

\section{MCMC diagnostics}\label{appE}

\begin{figure}[hbt!]
\begin{center}
\includegraphics[width=.8\textwidth]{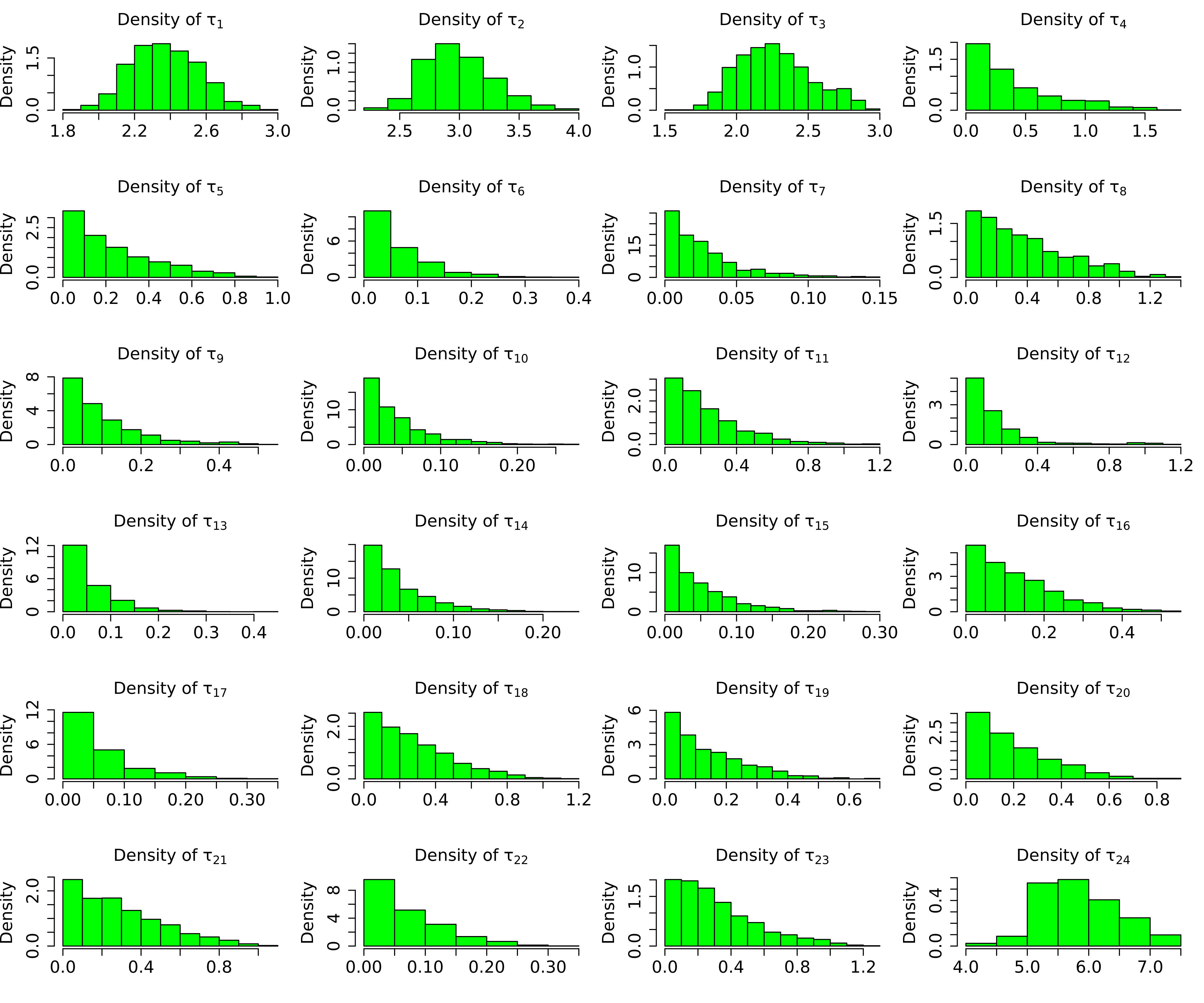}
\end{center}
\caption{Posterior summary showing the densities of each parameter for EU-wide data.}\label{hist_param}
\end{figure}

\begin{figure}[hbt!]
\begin{center}
\includegraphics[width=.8\textwidth]{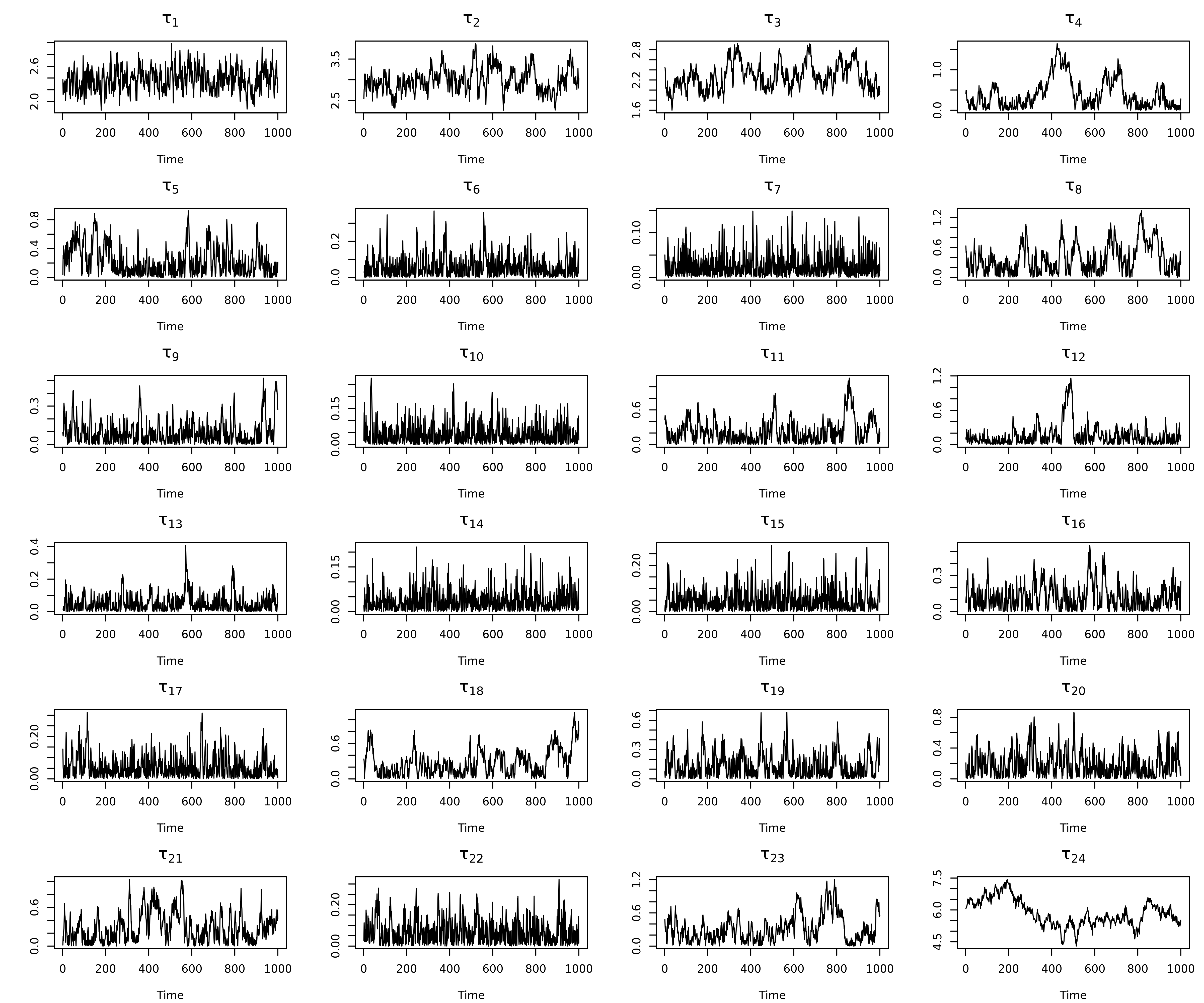}
\end{center}
\caption{Posterior summary showing the trace plots of each parameter for EU-wide data.}\label{trace_param}
\end{figure}

\begin{table}
\caption{Potential scale reduction factor for each MCMC chain, together with upper and lower confidence limits. Approximate convergence is diagnosed when the upper limit is close to 1.\label{est_param}}
\setlength{\tabcolsep}{3pt}
\centering
\begin{tabular}{rrr}
  \hline
 & Point est. & Upper C.I. \\
 \hline
1 & 1.01 & 1.01 \\
  2 & 1.05 & 1.09 \\
  3 & 1.00 & 1.01 \\
  4 & 1.82 & 3.49 \\
  5 & 1.04 & 1.17 \\
  6 & 1.02 & 1.06 \\
  7 & 1.00 & 1.01 \\
  8 & 1.04 & 1.18 \\
  9 & 1.01 & 1.01 \\
  10 & 1.00 & 1.01 \\
  11 & 1.33 & 2.04 \\
  12 & 1.17 & 1.58 \\
  13 & 1.00 & 1.00 \\
  14 & 1.02 & 1.02 \\
  15 & 1.00 & 1.01 \\
  16 & 1.02 & 1.09 \\
  17 & 1.00 & 1.02 \\
  18 & 1.20 & 1.67 \\
  19 & 1.00 & 1.02 \\
  20 & 1.06 & 1.23 \\
  21 & 1.16 & 1.57 \\
  22 & 1.01 & 1.05 \\
  23 & 1.06 & 1.15 \\
  24 & 1.10 & 1.39 \\
   \hline
\end{tabular}
\end{table}

\end{appendices}

\end{document}